\RequirePackage{ifpdf}
\ifpdf
\else
\PassOptionsToPackage{dvipdfmx}{graphicx}
\PassOptionsToPackage{dvipdfmx}{hyperref}
\fi

\documentclass[a4paper,11pt]{article}

\usepackage[dvipdfmx]{graphicx}
\usepackage{jheppub}
\usepackage[T1]{fontenc}

\usepackage{physics,color}

\usepackage[hang,small,bf]{caption}
\usepackage[subrefformat=parens]{subcaption}
\usepackage{bm}
\usepackage[normalem]{ulem}

\author[a]{Takaaki Ishii,}
\author[b]{Ryo Kitaku,}
\author[c]{Keiju Murata,}
\author[b]{Chul-Moon Yoo}

\affiliation[a]{Department of Physics, Rikkyo University, Nishi-Ikebukuro, Tokyo 171-8501, Japan}
\affiliation[b]{Division of Particle and Astrophysical Science, Graduate School of Science, Nagoya University, Furo-cho,
Chikusa-ku, Nagoya 464-8602, Japan}
\affiliation[c]{Department of Physics, College of Humanities and Sciences, Nihon University, Sakura-josui, Tokyo 156-8550, Japan}

\emailAdd{ishiitk@rikkyo.ac.jp}
\emailAdd{kitaku.ryo.f4@s.mail.nagoya-u.ac.jp}
\emailAdd{murata.keiju@nihon-u.ac.jp}
\emailAdd{yoo.chulmoon.k6@f.mail.nagoya-u.ac.jp}

\abstract{
We demonstrate the turbulent dynamics of the Nambu-Goto open string in the AdS$_3$ spacetime. While the motion of a classical closed string in AdS is known to be integrable, the integrability of an open string motion depends on the boundary conditions at the string endpoints. We numerically solve the equations of motion of the open string under the boundary conditions where the endpoints are i) fixed to a finite radial coordinate in AdS, and ii) free. For i), 
we 
find turbulence on the string, that shows a cascade in the energy and angular momentum spectra. This result indicates the non-integrability of the open string with this type of boundary conditions. For ii), we find no turbulence. This is consistent with the integrability of the open string with the free boundary conditions.
}

\title{
    Turbulence on open string worldsheets under non-integrable boundary conditions
}
\date{}
\preprint{RUP-23-23}
\begin{document}
\maketitle

\section{Introduction}
The AdS/CFT correspondence has received a great deal of attention since its discovery \cite{Maldacena:1997re}. One of the most typical examples is the duality between the type IIB superstring theory in AdS$_5\times S^5$ and the $\mathcal{N}=4$ $SU(N)$ 4D super Yang-Mills theory, especially in the classical and large $N$ limits. In testing this correspondence, integrability played a significant role. (See Ref.~\cite{Beisert:2010jr} and references therein for a comprehensive review.)

The string theory in AdS$_5\times S^5$ is known to be integrable, at least semiclassically ~\cite{Mandal:2002fs,Bena:2003wd}.
To be precise, this is the integrability of the theory with closed strings.
Let us restrict our attention to a classical closed string in AdS$_3 \subset$ AdS$_5\times S^5$ for simplicity. It is known that 
the string dynamics in AdS$_3$ can be described by the principal chiral model (PCM) whose target space is an $SL(2,R)$ group manifold. (See \cite{Yoshidabook} for a review.) When there are no boundaries in the PCM, that is the case for a closed string, an infinite number of conserved Yangian charges can be constructed, and this proves the integrability of the closed string in the AdS$_3$~\cite{Evans:1999mj}.

If an open string is considered, it is necessary to take into account the boundaries in the PCM. If the PCM has boundaries, there are non-zero fluxes of Yangian charges from the boundaries, and the integrability would be broken in general. However, it has been shown that, under certain boundary conditions, an infinite number of conserved charges can still be constructed and maintain integrability~\cite{MacKay:2001bh,Delius:2001he,Mann:2006rh,Dekel:2011ja,MacKay:2011zs}. 
An open string is not in the original type IIB superstring theory in AdS$_5\times S^5$, but it can be additionally introduced.
For example, a probe string hanging from the AdS boundary is considered to be dual to the Wilson loop operator giving rise to quark-antiquark potential in $\mathcal{N}=4$ super Yang-Mills theory \cite{Rey:1998ik,Maldacena:1998im}. Integrability is not obvious for such open strings.
It would be interesting to see under what conditions the integrability is maintained or not for an open string in AdS. 
Recently in Ref.~\cite{Ishii:2023ucw}, sufficient conditions for the integrability of a classical open string in AdS$_3$ were explicitly classified.
In this paper, we aim to understand how the motion of an open string in AdS$_3$ differs under such integrable and non-integrable boundary conditions.

Specifically, there are choices of Neumann and Dirichlet boundary conditions at the boundaries of the string worldsheet. The line element of AdS$_3$ can be given by
\begin{equation}
    ds^2=-\ell^2\left(\frac{1+r^2}{1-r^2}\right)^2dt^2+\frac{4\ell^2}{(1-r^2)^2}\left(dr^2+r^2d\theta^2\right) ,
    \label{AdS3 metric}
\end{equation}
where $\ell$ is the $\mathrm{AdS}_3$ radius and the AdS boundary is located at $r=1$. 
The embedding of an open string in AdS$_3$ is specified by three coordinate variables $(t,r,\theta)$ described as functions of worldsheet coordinates. 
On each coordinate variable, we can impose Neumann or Dirichlet boundary conditions on the boundaries of the worldsheet. 
For each boundary, there are $2^3=8$ possibilities for the choices of the boundary conditions. 
We will express these as $(\mathrm{N, N, N})$, $(\mathrm{N, N, D})$, $(\mathrm{N, D, N})$, and so on. 
This notation denotes the Neumann (N) or Dirichlet (D) condition for $(t,r,\theta)$ in this order.
For example, $(\mathrm{N, N, D})$ corresponds to imposing the Neumann, Neumann, and Dirichlet boundary conditions on $t, \, r$, and $\theta$ coordinates, respectively.

\begin{figure}[t]
    \centering
    \begin{minipage}[b]{0.45\linewidth}
        \centering
        \includegraphics[keepaspectratio, scale=0.4]{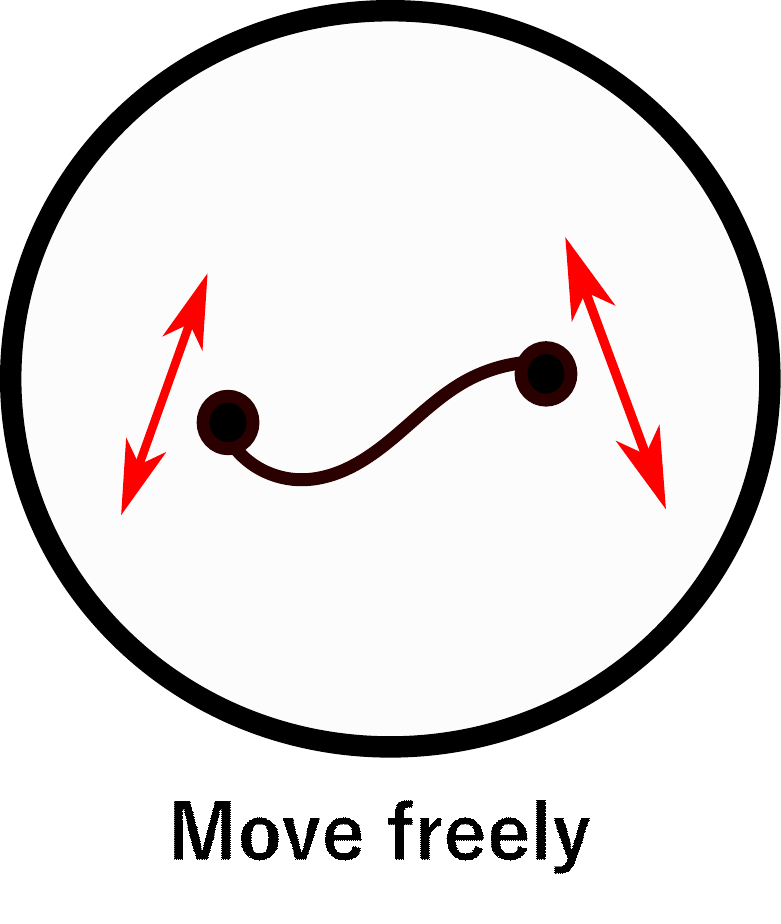}
        \subcaption*{$(\mathrm{N, N, N})$}
        \label{figure:boundarycondtionNNN}
        \end{minipage}
    \begin{minipage}[b]{0.45\linewidth}
    \centering
    \includegraphics[keepaspectratio, scale=0.4]{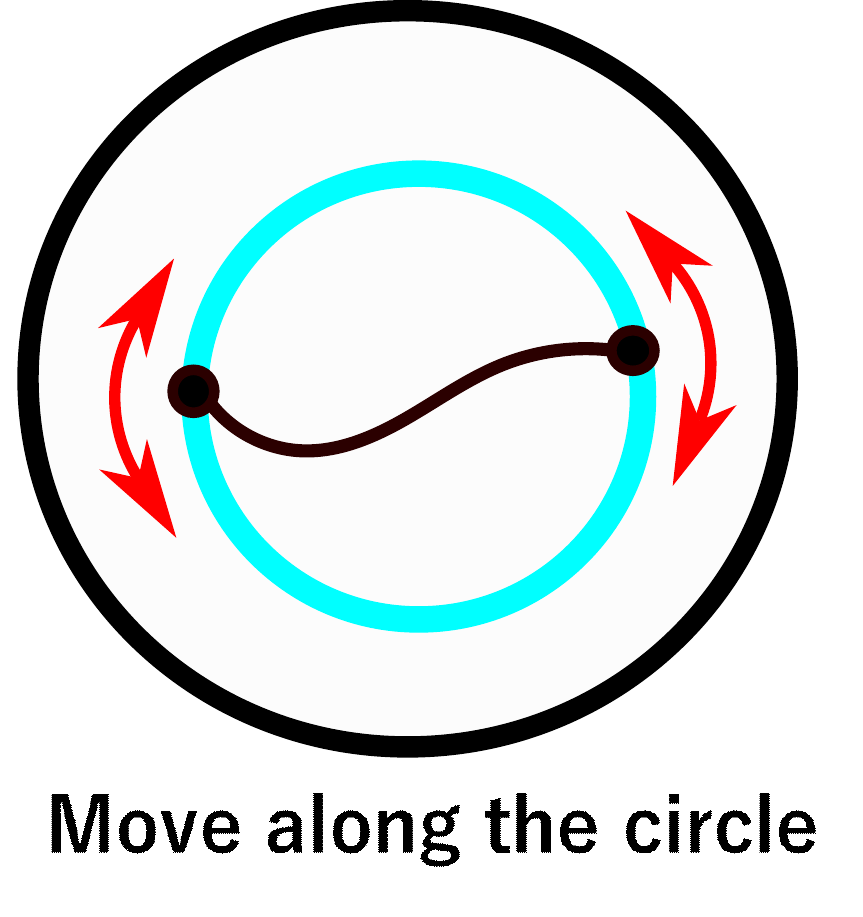}
    \subcaption*{$(\mathrm{N, D, N})$}
    \label{figure:boundarycondtionNDN}
    \end{minipage}    
    \caption{Schematic picture of an open string with two different boundary conditions. The black circle is the AdS boundary and inside is the AdS bulk. The open string is shown by the black curved line with the endpoints denoted by the black dots. In the right figure, the light blue circle is the place where the string endpoints are bounded. }
    \label{figure:stringmotionmodel}
\end{figure}

In this paper, we focus on the cases of $(\mathrm{N, N, N})$ and $(\mathrm{N, D, N})$.
Schematic pictures of such open strings in AdS$_3$ are given in figure~\ref{figure:stringmotionmodel}.
The case of $(\mathrm{N, N, N})$ corresponds to an open string in AdS with free endpoints, where the string endpoints are floating in the AdS bulk.  
For $(\mathrm{N, D, N})$, we can imagine that there is a ``D-brane'' (colored by light blue in figure \ref{figure:stringmotionmodel}) on a constant radius surface.
The endpoints of the string are bounded on the D-brane but can freely move in its tangential direction.

In particular, we consider the time evolution of an open string when a small perturbation is added to a reference steady rotating string given by the same configuration as the Gubser-Klebanov-Polyakov (GKP) string. The GKP string is a solidly rotating folded closed string in AdS, but it can be also regarded as a solidly rotating open string in AdS$_3$. Let us call it a GKP open string. The steady GKP open string does not distinguish $(\mathrm{N, N, N})$ and $(\mathrm{N, D, N})$ boundary conditions. However, when it is perturbed, the time evolution of the oscillating string will be different for these cases because of the boundary conditions imposed on the endpoints.

Here, we numerically study the time evolution of the perturbed GKP open string with the $(\mathrm{N, N, N})$ and $(\mathrm{N, D, N})$ boundary conditions and investigate their difference as claimed in~\cite{Ishii:2023ucw}.
For $(\mathrm{N, N, N})$, open strings are integrable~\cite{Mann:2006rh,Dekel:2011ja}. 
Therefore, we expect the motion to be quasiperiodic in analogy with finite-dimensional dynamical systems. 
For $(\mathrm{N, D, N})$, an infinite number of conserved charges cannot be constructed in the same way as $(\mathrm{N, N, N})$. 
Then, we will see turbulence in the dynamics of an open string with $(\mathrm{N, D, N})$. Such a result will strongly suggest non-integrability of the system. 

The non-integrable motion of a string was also studied in asymptotically  AdS spacetime.\footnote{An analytic approach of non-integrability can be found in Ref.~\cite{Basu:2011fw,Rigatos:2020igd}.}
For a closed string, one can find chaotic motion \cite{PandoZayas:2010xpn,Basu:2011dg,Basu:2011di,Asano:2015qwa,Ishii:2016rlk,Kushiro:2022ksg,Panigrahi:2016zny,Ishii:2021asw} and 
turbulence \cite{Ishii:2016rlk}.
For an open string, chaotic behavior and turbulence also have been observed \cite{Hashimoto:2018fkb,Akutagawa:2019awh}.
The geometries in these works are not the pure AdS spacetime where the motion of a closed string is integrable.
However, in Ref.~\cite{Ishii:2015wua,Ishii:2015qmj,Ishii:2023ucw}, even in the pure AdS spacetime, 
turbulence was found in the motion of the open string whose endpoints are fixed on the AdS boundary.
For the emergence of the turbulence, the relevance of boundary conditions on the string was argued \cite{Ishii:2023ucw}. 
 Our result will give another example of the turbulence of the open string in AdS driven by the non-integrable boundary conditions.

This paper is organized as follows. In section \ref{section:AdSstringmotion}, 
we derive the equation of motion of the Nambu-Goto string, and 
define the energy and angular momentum for the string motion. 
In section \ref{section:GKPstring}, we explain the setting for the string dynamics. 
The GKP string is introduced as a reference background solution, and a small perturbation is applied on top of it.
Numerical results are shown in section \ref{section:Result}, where 
subsections \ref{subsection:initial condition} and \ref{subsection:turbulance} 
are devoted to the results of the string motion and turbulence, respectively. 
We conclude in section \ref{section:conclusion}.
Details of the numerical scheme and the estimation of numerical errors are displayed in appendices \ref{appendix:numericalsheme} and \ref{appendix:Error analysis}. 
In Appendix \ref{appendix:conserved charge}, we explicitly show that the conserved charges for $(\mathrm{N, N, N})$ are not conserved for $(\mathrm{N, D, N})$. 
In Appendix \ref{appendix:chaos}, we discuss the results of our analysis of the sensitivity to initial conditions for $( \mathrm{N, D, N})$. 
In Appendix \ref{appendix:numerical check of E J}, we check the conservations of energy and angular momentum for $(\mathrm{N, D, N})$ numerically.

\section{Nambu-Goto strings in AdS$_3$ spacetime}
\label{section:AdSstringmotion}
\subsection{Equations of motion}
\label{subsection:AdS spacetime}
For convenience in numerical calculations, we introduce 
``Cartesian'' coordinates $\boldsymbol{\chi}=(\chi_1,\chi_2)$ (see also Ref.~\cite{Ishii:2015qmj}) as 
\begin{equation}
    \chi_1=r \cos\theta,\quad \chi_2=r\sin \theta. 
    \label{poloarcoordinate}
\end{equation}
They are defined in $|\boldsymbol{\chi}|<1$.
Then, the AdS$_3$ metric \eqref{AdS3 metric} is expressed as 
\begin{equation}
    ds^2= -\ell^2\left(\frac{1+|\boldsymbol{\chi}|^2}{1-|\boldsymbol{\chi}|^2}\right)^2dt^2+\frac{4\ell^2}{(1-|\boldsymbol{\chi}|^2)^2}d\boldsymbol{\chi}\cdot d\boldsymbol{\chi}. 
    \label{AdS3 metric cartesian}
\end{equation}
A motivation of introducing these coordinates is that the polar coordinates $(r,\theta)$ are singular at $r=0$, but the metric is explicitly regular at $\boldsymbol{\chi}=0$ in the Cartesian coordinates.
In the rest of the paper, we set $\ell=1$.

We consider the string motion described by the Nambu-Goto action, 
\begin{equation}
    S_{\mathrm{NG}}=-\frac{1}{2\pi\alpha'}\int d^2\sigma \sqrt{-h},\quad h\equiv \det(h_{ab}), 
    \label{NGaction}
\end{equation}
with $h_{ab}$ being the induced metric given by 
\begin{equation}
    h_{ab}=g_{\alpha\beta}x^{\alpha}_{,a}x^{\beta}_{,b},  
    \label{induced metric}
\end{equation}
where the Greek and Roman indices label the coordinates on the target spacetime and the string worldsheet, respectively.  
Let $(u,v)$ denote the worldsheet coordinates.
Then, the components of the 
induced metric \eqref{induced metric} in terms of the spacetime coordinates $(t,\chi_1,\chi_2)$ are expressed as 
\begin{equation}
    \begin{aligned}
    &h_{uu}=-\left(\frac{1+|\boldsymbol{\chi}|^2}{1-|\boldsymbol{\chi}|^2}\right)^2t_{,u}^2+\frac{4}{(1-|\boldsymbol{\chi}|^2)^2}|\boldsymbol{\chi}_{,u}|^2, \\
    &h_{vv}=-\left(\frac{1+|\boldsymbol{\chi}|^2}{1-|\boldsymbol{\chi}|^2}\right)^2t_{,v}^2+\frac{4}{(1-|\boldsymbol{\chi}|^2)^2}|\boldsymbol{\chi}_{,v}|^2, \\
    &h_{uv}=-\left(\frac{1+|\boldsymbol{\chi}|^2}{1-|\boldsymbol{\chi}|^2}\right)^2t_{,v}t_{,u}+\frac{4}{(1-|\boldsymbol{\chi}|^2)^2}\boldsymbol{\chi}_{,u}\cdot \boldsymbol{\chi}_{,v}. 
    \end{aligned}
    \label{induced metric explicit}
\end{equation} 

Since the worldsheet is a two-dimensional surface, 
the induced metric can be rewritten in the conformally flat form with an appropriate set of the 
worldsheet coordinates. 
Here, we impose 
\begin{equation}
    h_{uu}=h_{vv}=0\label{Virasoro}.
\end{equation}
Under these conditions, the coordinates $(u,v)$ are called double null coordinates.  
Using the double null coordinates, we find $\sqrt{-h}=\sqrt{h_{uv}^2-h_{uu}h_{vv}}=-h_{uv}$ in the Nambu-Goto action~(\ref{NGaction}). (Note that $h_{uv}<0$ since we choose both $\partial _u$ and $\partial _v$ as future-directed vectors.) 
Then, the Nambu-Goto action \eqref{NGaction} becomes 
\begin{equation}
    S= \frac{1}{2\pi\alpha'}\int dudv \left(-\left(\frac{1+|\boldsymbol{\chi}|^2}{1-|\boldsymbol{\chi}|^2}\right)^2t_{,v}t_{,u}+\frac{4}{(1-|\boldsymbol{\chi}|^2)^2}\boldsymbol{\chi}_{,u}\cdot \boldsymbol{\chi}_{,v}\right).
    \label{Virasoroaction}
\end{equation}
The equations of motion can be derived from the above action 
as
\begin{equation}
 \begin{aligned}
    &t_{,uv}=\frac{-4}{(1-|\boldsymbol{\chi}|^2)(1+|\boldsymbol{\chi}|^2)}\left((\boldsymbol{\chi}_u\cdot \boldsymbol{\chi})t_{,v}+(\boldsymbol{\chi}_v\cdot \boldsymbol{\chi})t_{,u}\right), \\
    &\boldsymbol{\chi}_{,uv}=-\frac{1}{(1-|\boldsymbol{\chi}|^2)}\big(2(\boldsymbol{\chi}\cdot\boldsymbol{\chi}_{,u})\boldsymbol{\chi}_{,v}\\
&\hspace{3cm}
-2(\boldsymbol{\chi}_{,u}\cdot\boldsymbol{\chi}_v)\boldsymbol{\chi}+2(\boldsymbol{\chi}\cdot\boldsymbol{\chi}_{,v})\boldsymbol{\chi}_{,u}+(1+|\boldsymbol{\chi}|^2)t_{,u}t_{,v}\boldsymbol{\chi}\big) .
    \label{stringeom}
 \end{aligned}
\end{equation}

To solve the equations of motion, it is crucial to make sure that the double null constraints \eqref{Virasoro} are imposed. First,
    as usual constraint systems, if the constraints \eqref{Virasoro} are satisfied on the initial surface and boundaries, 
they are guaranteed to be satisfied during time evolution. 
    Hence, we can use the constraints \eqref{Virasoro} to check the accuracy of numerical calculations. (See Appendix~\ref{appendix:Error analysis}.)
Second,
    since $\partial_u$ and $\partial_v$ should be future-directed vectors, i.e.~$t_{,u}>0$ and $t_{,v}>0$, 
   we can solve the constraint equations \eqref{Virasoro} as
\begin{equation}
    t_{,u}=\frac{2}{(1+|\boldsymbol{\chi}|^2)}|\boldsymbol{\chi}_{,u}|,\quad t_{,v}=\frac{2}{(1+|\boldsymbol{\chi}|^2)}|\boldsymbol{\chi}_{,v}|. 
    \label{constraintstutv}
\end{equation}
In order to realize stable numerical integration~\cite{Ishii:2015qmj,Ishii:2015wua}, by using (\ref{constraintstutv}), we eliminate $t_{,u}$ and $t_{,v}$ from evolution equations~(\ref{stringeom}) as
\begin{equation}
    \begin{aligned}
        &t_{,uv}=\frac{-8}{(1-|\boldsymbol{\chi}|^2)(1+|\boldsymbol{\chi}|^2)^2}
        \left((\boldsymbol{\chi}_u\cdot \boldsymbol{\chi})|\boldsymbol{\chi}_{,v}|
        +(\boldsymbol{\chi}_v\cdot \boldsymbol{\chi})|\boldsymbol{\chi}_{,u}|\right), \\
        &\boldsymbol{\chi}_{,uv}=-\frac{2}{1-|\boldsymbol{\chi}|^4}\big(
        2|\boldsymbol{\chi}_{,u}||\boldsymbol{\chi}_{,v}|\boldsymbol{\chi}+\\
        &\hspace{3cm}(1+|\boldsymbol{\chi}|^2)((\boldsymbol{\chi}\cdot\boldsymbol{\chi}_{,u})
        \boldsymbol{\chi}_{,v}+(\boldsymbol{\chi}\cdot\boldsymbol{\chi}_{,v})\boldsymbol{\chi}_{,u}
        -(\boldsymbol{\chi}_{,u}\cdot\boldsymbol{\chi}_v)\boldsymbol{\chi})\big).
    \end{aligned}
    \label{stringeom2} 
\end{equation}

We still have the residual gauge degrees of freedom associated with the coordinate transformations 
from $u$ and $v$ to arbitrary functions of $u$ and $v$, respectively. 
By using these degrees of freedom, we can fix the range of the worldsheet coordinate as $-\pi/2 \le u-v \le \pi/2$.
We introduce another coordinate system $(\tau,\sigma)$ as 
\begin{equation}
    \tau=u+v,\quad \sigma=u-v. 
    \label{spatial,tau}
\end{equation}
The boundaries of the string worldsheet are located at $\sigma=u-v=-\pi/2,\pi/2$. 

For an open string, we need to impose boundary conditions on the worldsheet boundaries. In this paper, we consider two types of boundary conditions: $(\mathrm{N, N, N})$ and $(\mathrm{N, D, N})$.
The $(\mathrm{N, N, N})$ boundary conditions are given by
 \begin{equation}
    \partial_{\sigma}t\left(\tau,\sigma=\pm\frac{\pi}{2}\right)=0,\quad \partial_{\sigma}r\left(\tau,\sigma=\pm\frac{\pi}{2}\right)=0,\quad \partial_{\sigma}\theta\left(\tau,\sigma=\pm\frac{\pi}{2}\right)=0.
    \label{boundarycondtionNNN}
    \end{equation}
The $(\mathrm{N, D, N})$ boundary conditions can be written as 
\begin{equation}
 \partial_{\sigma}t\left(\tau,\sigma=\pm\frac{\pi}{2}\right)=0,\quad r\left(\tau,\sigma=\pm\frac{\pi}{2}\right)=r_0=\textrm{const.},\quad \partial_{\sigma}\theta\left(\tau,\sigma=\pm\frac{\pi}{2}\right)=0,   
 \label{boundarycondtionNDN}
 \end{equation}
where $r_0$ corresponds to the coordinate that the string endpoints are fixed in the AdS target space.
For technical details of the numerical evolution in the bulk and at the boundaries under these conditions, see appendices~\ref{appendix:numericalmethod}~and~\ref{appendix:boundary_time_evolution}.

\subsection{Energy and angular momentum}
\label{subsection:conserved}
We can evaluate the energy $E$ and angular momentum $J$ as conserved quantities in the time evolution of the string. 
In the $(\tau,\sigma)$ coordinates (\ref{spatial,tau}), the Nambu-Goto action under the double null constraints (\ref{Virasoroaction})
becomes
\begin{equation}
    \begin{aligned}
        S&=\frac{1}{4\pi\alpha'}\int d\tau d\sigma\left(-\left(\frac{1+r^2}{1-r^2}\right)^2
        (t_{,\tau}t_{,\tau}-t_{,\sigma}t_{,\sigma})\right.\\
        &\hspace{4cm} \left.+\frac{4}{(1-r^2)^2}((r_{,\tau}r_{,\tau}-r_{,\sigma}r_{,\sigma})
        +r^2(\theta_{,\tau}\theta_{,\tau}-\theta_{,\sigma}\theta_{,\sigma}))\right), \label{Virasoroactionpolar}
\end{aligned}
\end{equation}
where we used the polar coordinates $(r,\theta)$ instead of $(\chi_1,\chi_2)$ so that the symmetry of the system is manifest.
Since the above action is invariant under the time translation $t\to t+\textrm{const.}$ and rotation $\theta\to \theta+\textrm{const.}$, we can define conserved energy and angular momentum.
The conjugate momenta of $t$ and $\theta$, denoted by $p_t$ and $p_{\theta}$, can be given by
\begin{equation}
    p_t=\left(\frac{1+r^2}{1-r^2}\right)^2t_{,\tau},\quad p_{\theta}=\left(\frac{r}{1-r^2}\right)^2\theta_{,\tau}, 
    \label{pt,ptheta}
\end{equation}
where we have omitted the constant factor associated with the string tension for notational simplicity. 
Then, we define the energy $E$ and angular momentum $J$ as 
\begin{equation}
    E=\left(\frac{2}{\pi}\right)^2\int_{-\pi/2}^{\pi/2}d\sigma p_t,\quad  J=\left(\frac{2}{\pi}\right)^2\int_{-\pi/2}^{\pi/2}d\sigma p_\theta.
    \label{conservedchargeEJ}
\end{equation}
The factor $(2/\pi)^2$ in \eqref{conservedchargeEJ} is introduced for simplicity when we will define the Fourier coefficients in subsection \ref{subsection:turbulance}.
As a consequence of the transitional symmetry of $t$ and $\theta$, 
the time derivative of $E$ and $J$ are given by boundary terms as
\begin{equation}
    \dv{E}{\tau}=\left(\frac{2}{\pi}\right)^2 \left[\left(\frac{1+r^2}{1-r^2}\right)^2t_{,\sigma}\right]^{\pi/2}_{-\pi/2}, \quad \dv{J}{\tau}=\left(\frac{2}{\pi}\right)^2 \left[\left(\frac{r}{1-r^2}\right)^2\theta_{,\sigma}\right]^{\pi/2}_{-\pi/2}. 
    \label{E,theta,dt}
\end{equation}
Since  we consider only the Neumann boundary conditions for $t$ and $\theta$ coordinates, i.e.~$t_{,\sigma}|_{\sigma=\pm \frac{\pi}{2}}=\theta_{,\sigma}|_{\sigma=\pm \frac{\pi}{2}}=0$, 
$E$ and  $J$ are conserved. 
In terms of the coordinates $(t,\boldsymbol{\chi})$, the conjugate momenta are 
written as 
\begin{equation}
\begin{split}
    &p_t=\left(\frac{1+|\boldsymbol{\chi}|^2}{1-|\boldsymbol{\chi}|^2}\right)^2t_{,\tau}, \\
    &p_\theta=\boldsymbol{\chi}\times \frac{\boldsymbol{\chi}_{,\tau}}{(1-|\boldsymbol{\chi}|^2)^2}=\boldsymbol{\chi}\times p_{\boldsymbol{\chi}}, 
\end{split}
    \label{ptptheta,cartesian}
\end{equation}
where $p_{\boldsymbol{\chi}}$ is the conjugate momenta of $\boldsymbol{\chi}$. 

\section{Setup of the numerical simulation}
\label{section:GKPstring}

First, we introduce the Gubser-Klebanov-Polyakov (GKP) string, which is an exact solution on AdS$_3$ spacetime~\cite{Gubser:2002tv}. We use this as a reference solution and we will add perturbations on the GKP string.
In our coordinate system, the solution can be explicitly parametrized by the worldsheet coordinates as 
\begin{equation}
\begin{split}
    &t^{\mathrm{GKP}}(\tau,\sigma)=\kappa \tau, \\
 &\chi^{\mathrm{GKP}}_1(\tau,\sigma)+i\chi^{\mathrm{GKP}}_2(\tau,\sigma)=\frac{\sqrt{1-k^2}-\mathrm{dn}(\omega(k) \sigma +\boldsymbol{K}(k)|k^2)}{k\ \mathrm{cn}(\omega(k) \sigma +\boldsymbol{K}(k)|k^2)}e^{i\omega(k) \tau},
\end{split}
    \label{GKPstring}
\end{equation}
where $\omega(k)=2 \boldsymbol{K}(k)/\pi$ and $\kappa=k\,\omega(k)$ is a constant parameter. Here, $\mathrm{cn}(x|k^2)$ and $\mathrm{dn}(x|k^2)$ are the Jacobi elliptic functions with $\boldsymbol{K}$ being a complete elliptic integral of the first kind with the parameter $k$:
\begin{equation}
    \boldsymbol{K}(k)=\int_{0}^1 dt \frac{1}{\sqrt{(1-t^2)(1-k^2t^2)}}. 
    \label{complite integral of first kind}
\end{equation}

\begin{figure}[t]
    \centering
    \includegraphics[keepaspectratio,scale=0.4]{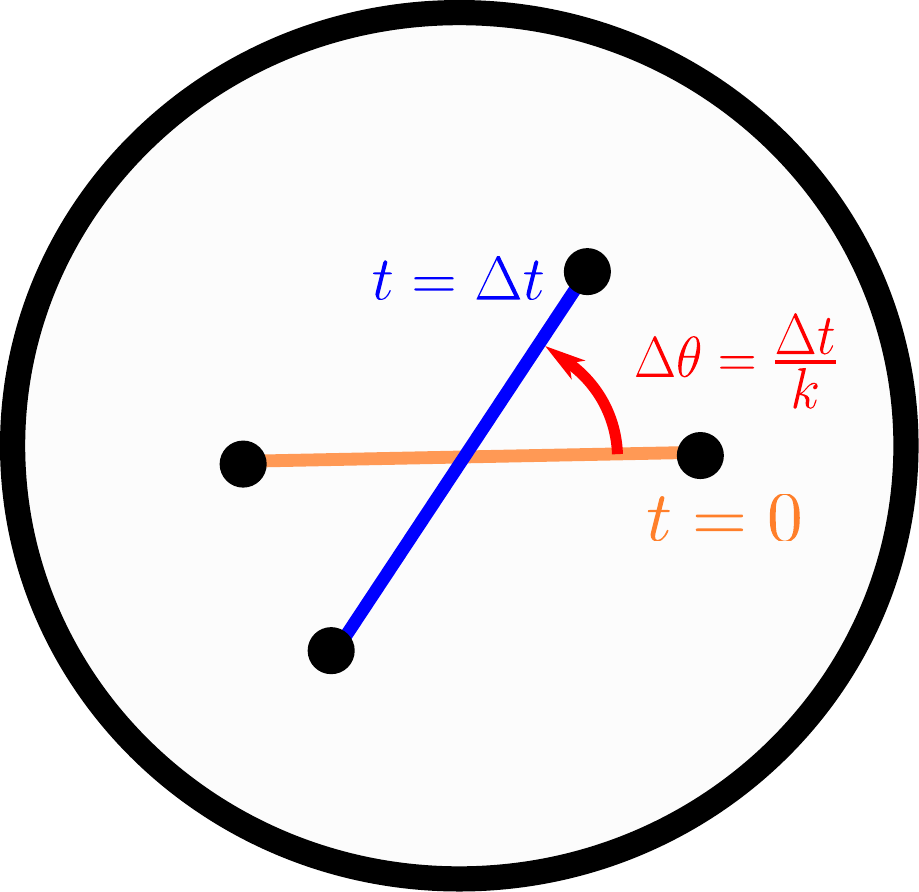}
    \caption{GKP string. The black circle is the AdS boundary.}
    \label{figure:GKP_string}
\end{figure}

The GKP string solution is expressed in the one-parameter family of $k$. 
This describes a rotating rod around the origin (see figure~\ref{figure:GKP_string}). 
The GKP string was originally introduced as a folded closed string, where in fact is given by taking the coordinate range as $-\pi\leq \sigma < \pi$ in (\ref{GKPstring}) with a periodic boundary condition.
In this paper, taking $-\pi/2\leq \sigma \leq \pi/2$, we regard it as the rigidly rotating open string.
The GKP solution satisfies both boundary conditions $(\mathrm{N, D, N})$ and $(\mathrm{N, N, N})$ defined in Eqs.(\ref{boundarycondtionNDN}) and (\ref{boundarycondtionNNN}). That is, the steady GKP solution (\ref{GKPstring}) does not distinguish the two boundary conditions. These make a difference when perturbation is added.

For the numerical simulation, we consider the following initial condition at $\tau=0$: 
\begin{equation}
    t(0,\sigma)=0,\quad \boldsymbol{\chi}(0,\sigma)=\boldsymbol{\chi}^{GKP}(0,\sigma),\quad \partial_{\tau}{\chi_2}(0,\sigma)=\partial_{\tau}{\chi_2^{\mathrm{GKP}}}(0,\sigma)+\epsilon \exp(-\tan^2\sigma), 
    \label{initial value}
\end{equation}
where $\epsilon$ is a small constant number. 
The value of $r_0$ for the reference GKP solution is given by 
$r_0=\chi_1^{\mathrm{GKP}}(0,\frac{\pi}{2})$. 
The other initial conditions $\partial_{\tau}{t}(0,\sigma)$ and $\partial_{\tau}{\chi_1^{\mathrm{GKP}}}(0,\sigma)$,
necessary for solving the second-order differential equations, are determined by the constraints \eqref{Virasoro}. See appendix \ref{appendix:constructionginitialvalue} for details.
After this section, we take $k$ as $k=0.5$.

\section{Results}
\label{section:Result}
\subsection{String motion}
\label{subsection:initial condition}

\begin{figure}[t]
  \begin{minipage}[b]{0.5\linewidth}
    \centering
    \includegraphics[keepaspectratio, scale=0.5]{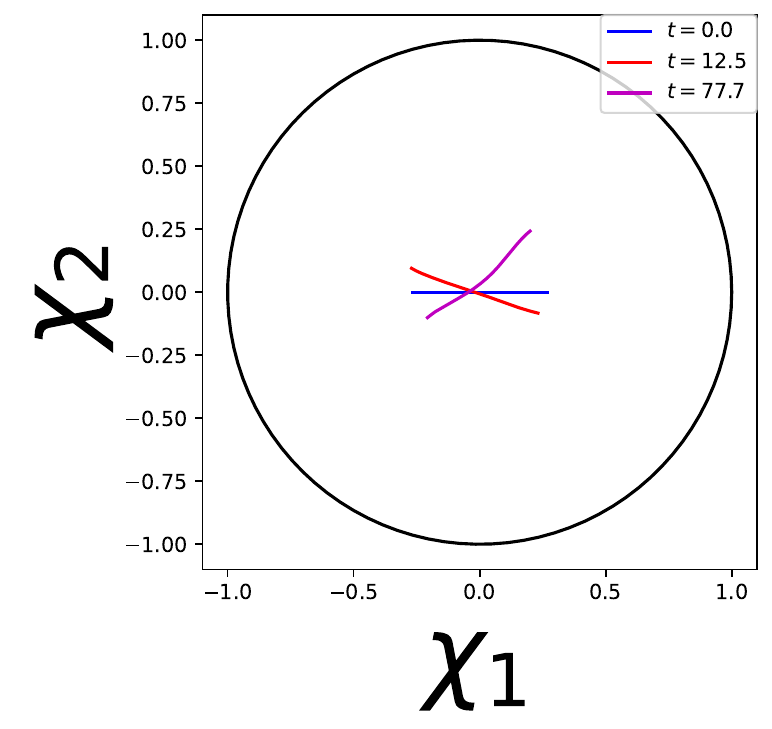}
    \subcaption{$ (\mathrm{N, N, N})$: $0.0\le t\le 77.7$}
    \label{figure:stringmotionNNN}
\end{minipage}
\begin{minipage}[b]{0.5\linewidth}
        \centering
        \includegraphics[keepaspectratio, scale=0.5]{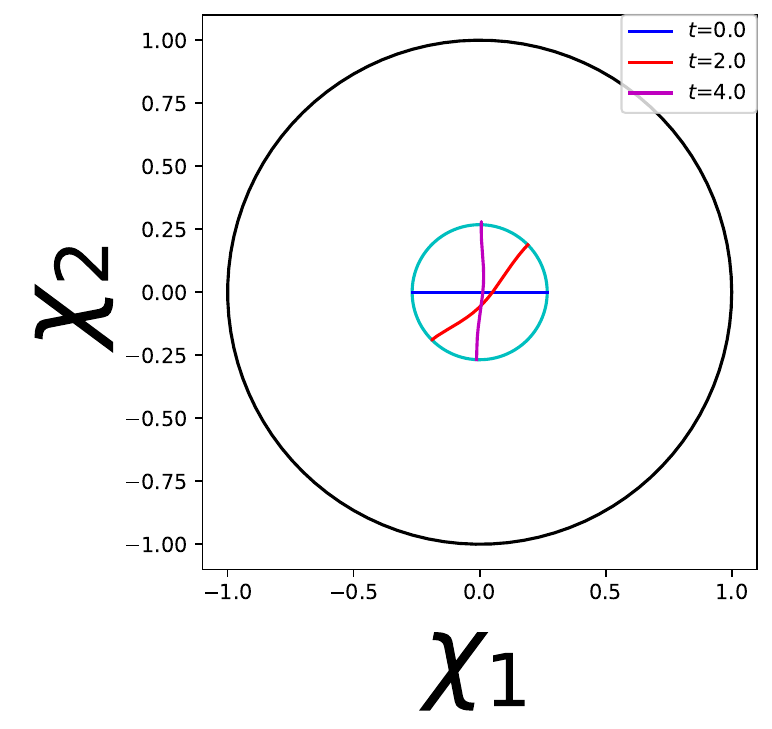}
        \subcaption{$ (\mathrm{N, D, N})$: $0.0\le t\le 4.0$}
        \label{figure:stringmotionNDNa}
    \end{minipage}\\[5mm]
    \begin{minipage}[b]{0.5\linewidth}
        \centering
        \includegraphics[keepaspectratio, scale=0.5]{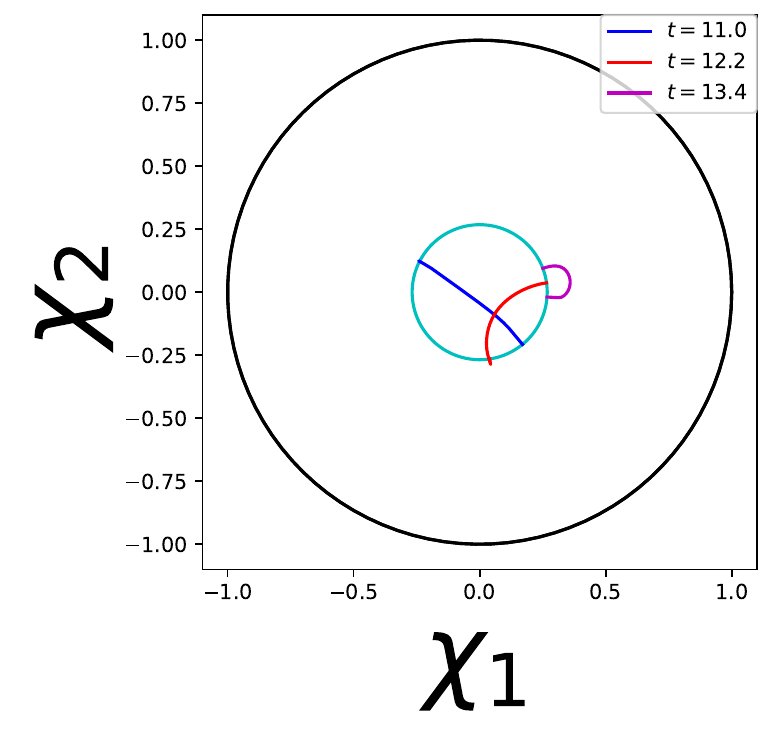}
        \subcaption{$ (\mathrm{N, D, N})$: $11.0\le t \le 13.4$}
        \label{figure:stringmotionNDNb}
    \end{minipage}
    \begin{minipage}[b]{0.5\linewidth}
        \centering
        \includegraphics[keepaspectratio, scale=0.5]{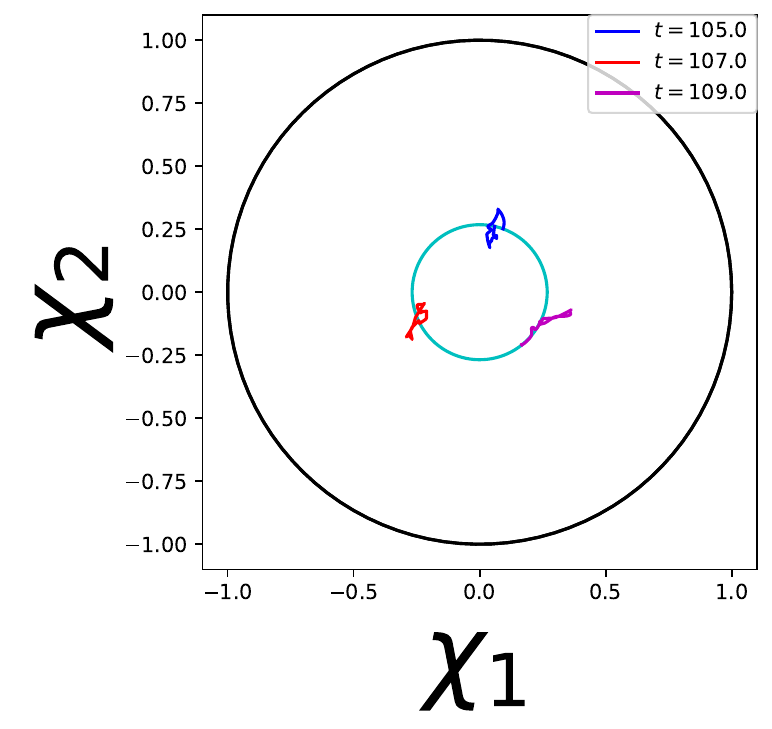}
        \subcaption{$ (\mathrm{N, D, N})$: $105.0\le t \le 109.0$}
        \label{figure:stringmotionNDNc}
    \end{minipage}
    \caption{Snapshots of the string motion. 
    Colored lines describe the perturbed string at different times. The black circle is the AdS boundary.  
    The light blue circle is the ``D-brane'' where the string endpoints are bounded.
    Panel (a) shows the string motion for $(\mathrm{N, N, N})$, and 
    the other panels are for $(\mathrm{N, D, N})$. 
    }
    \label{figure:motion of string}
\end{figure}

We first discuss the overall picture of the string motion. 
In this subsection, we take $\epsilon$ as $\epsilon=0.1$. 
Figure~\ref{figure:stringmotionNNN} shows snapshots of the string motion for $(\mathrm{N, N, N})$.
The string is kept stretched and not bent much, that is, the configuration is kept close to the straight line and not so much different from the reference GKP string.
Figures~\ref{figure:stringmotionNDNa}--\ref{figure:stringmotionNDNc} show snapshots of the string motion for $(\mathrm{N, D, N})$.
Initially, in $0.0\le t \le 4.0$ (figure~\ref{figure:stringmotionNDNa}), 
the string configuration looks similar to the reference GKP string.
However, in $11.0\le t \le 13.4$ (figure~\ref{figure:stringmotionNDNb}), 
we can start to see significant differences. 
Around this time, the perturbed string starts to repeat the motion of shrinking and stretching, unlike the case of $(\mathrm{N, N, N})$. The string configuration is different from the reference GKP string significantly. 
As we can see in $105.0\le t \le 109.0$ (figure~\ref{figure:stringmotionNDNc}), eventually, 
the perturbed string is crumpled into a small region and revolves along the D-brane. 
An intuitive interpretation of this behavior for $(\mathrm{N,D,N})$ is given by the shrinking of the string by its tension. 
For a GKP string, the tension and centrifugal force are balanced. 
For a perturbed $ (\mathrm{N,N,N})$ string, this does not seem to be unbalanced because the endpoints are free. However, for $(\mathrm{N,D,N})$, the string gradually slips off the antipodal points of the circle on which the endpoints are bounded, and when the string is away from the center, the tension wins and the string starts to shrink suddenly as we see in figure~\ref{figure:stringmotionNDNb}.
A more specific explanation of this change in motion for $ (\mathrm{N,D,N})$ is that the initial spin angular momentum of the string is transferred into the orbital angular momentum. (We will also see this behavior in the angular momentum spectrum in figure~\ref{figure:angularNDN}.)
As a result, for $(\mathrm{N,D,N})$, we can find an irregular motion. 
The qualitative difference of the crumpled string motion for $ (\mathrm{N,D,N})$  from $ (\mathrm{N,N,N})$
 might reflect the non-integrability due to the boundary conditions (see also Appendix~\ref{appendix:conserved charge}). 
In the following subsection, we will analyze the turbulence of the open string.\footnote{

        One might also expect that the one-dimensional angular motion of a string endpoint 
        would exhibit a chaotic character. However,  
        although we could observe an apparently random motion in the dynamics of the endpoint, 
        we did not find any clear chaotic character in the endpoint motion, 
        namely, we did not obtain strong evidences to conclude the presence of the sensitivity to initial conditions
        (see appendix \ref{appendix:chaos}).
}

\subsection{Turbulence}
\label{subsection:turbulance}
\subsubsection{Energy spectrum}
\label{turbulance:energy}
Turbulence can be characterized by the energy transfer between different modes in the 
energy spectrum caused by the non-linearity of the system.
For usual fluid mechanics, there is a cut-off scale below which the energy cascade is suppressed.
In the dynamics on asymptotically AdS spacetime, we often see turbulence~\cite{Bizon:2011gg, Bizon:2013xha,Ishii:2016rlk}.
These systems do not have dissipation unlike usual fluid mechanics.
Therefore arbitrary higher modes can be excited due to the energy cascade.

In order to check these behaviors, first, we need to appropriately 
define the energy spectrum. We want to define the energy spectrum by Fourier transformation.

However, in the current setting, since the boundary condition is not periodic, 
we suffer from the fictitious power-law spectrum associated with the sharp cutoff of the embedding function at the boundary. 
To avoid this fictitious power-law behavior, 
we introduce a new worldsheet coordinate $\sigma'$ given by
\begin{equation}
    \sigma=\frac{\pi}{2}\tanh(\tan \sigma ').
    \label{hensuuhenkan}
\end{equation}
Then the expression of $E$ \eqref{conservedchargeEJ} can be rewritten as\footnote{In Appendix \ref{appendix:numerical check of E J}, we check numerically that the energy $E$ and angular momentum $J$ are conserved in time evolution.}

\begin{equation}
 E=\frac{2}{\pi} \int _{-\pi/2}^{\pi/2} d\sigma' \frac{p_t(\sigma(\sigma'))}{\cosh^2(\tan(\sigma'))\cos^2(\sigma')}\equiv\frac{2}{\pi} \int_{-\pi/2}^{\pi/2}d\sigma' p'_t(\sigma'),   
\label{Esigmap} 
\end{equation}
where 
\begin{equation}
  p'_t(\sigma')=\frac{p_t(\sigma(\sigma'))}{\cosh^2(\tan(\sigma'))\cos^2(\sigma')}. 
\label{pprime}
 \end{equation} 
Although $p_t$ can have finite values at endpoints of the string $\sigma=\pm \pi/2$, 
$p_t'$ approaches zero exponentially at endpoints $\sigma'\to \pm \pi/2$. 

Considering the Fourier transformation\footnote{Because $\sqrt{p'_{t}}$ is a smooth function of $\sigma'$, the Fourier coefficients decay faster than any power functions of $n$ in $n \to \infty$.} 
of $\sqrt{p'_t}$ as

\begin{eqnarray}
    \begin{aligned}
    \sqrt{p'_t}&=\frac{C_0}{\sqrt{2}}+\frac{1}{2}\sum_{n=1}^\infty( C_ne^{2in\sigma'}+ C_{-n}e^{-2in\sigma'})\\
    &=\frac{C_0}{\sqrt{2}}+\frac{1}{2}\sum_{n=1}^\infty(C_ne^{2in\sigma'}+C^*_ne^{-2in\sigma'}), 
\end{aligned}
\label{modeenergy}
\end{eqnarray}
we obtain 
\begin{equation}
  E=\sum_{n=0}^\infty E_n
\label{EEn}
\end{equation}
with 
\begin{equation}
    E_n=|C_n|^2 .
    \label{suumatioin_of_modeenergy} 
\end{equation}
Although the energy spectrum depends on the choice of coordinates, we shall use fixed special coordinates $(\tau,\sigma')$ and study the qualitative behavior of the time dependence of the energy spectrum.

\begin{figure}[t]
    \centering
    \begin{minipage}[b]{0.45\linewidth}
      \centering
      \includegraphics[keepaspectratio, scale=0.4]{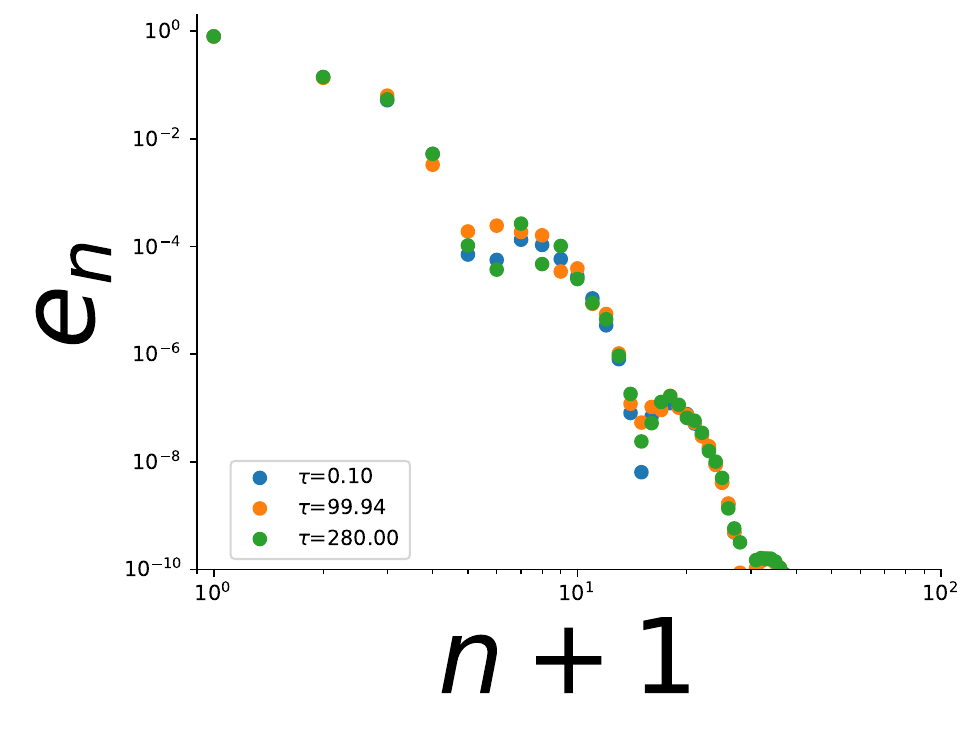}
      \subcaption{$(\mathrm{N, N, N})$}
      \label{figure:energyNNN}
      \end{minipage}
  \begin{minipage}[b]{0.45\linewidth}
    \centering
    \includegraphics[keepaspectratio, scale=0.4]{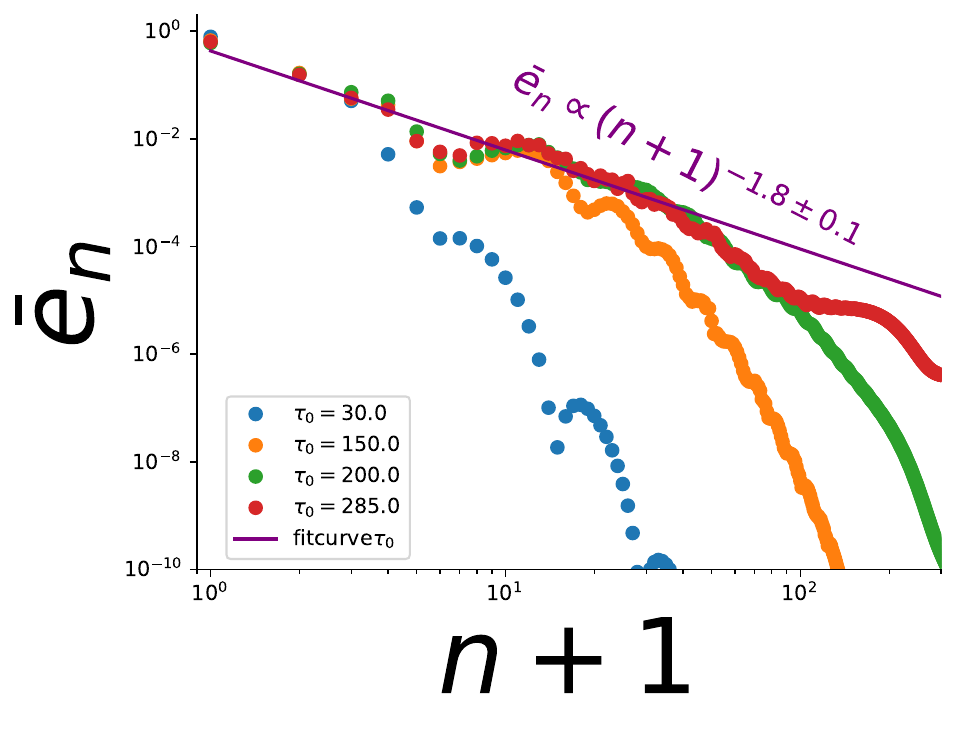}
    \subcaption{$(\mathrm{N, D, N})$}
    \label{figure:energyNDN}
    \end{minipage}
    \caption{
    Energy spectra for $(\mathrm{N, N, N})$ (panel (a)) and $(\mathrm{N,D,N})$ (panel (b)) at each given time. 
    In panel (b), the purple line is a fitting curve for the red points in the range $1\le n+1\le 40$.}
    \label{figure:energy}
\end{figure}

Figure~\ref{figure:energy} shows the energy spectra for several values of $\tau$. 
For $(\mathrm{N, N, N})$ (figure~\ref{figure:energyNNN}, where $ e_n\equiv E_n/E$), the energy spectra do not change in time so much.
That is, there is no turbulence on the worldsheet.
This explains why the motion of the string for $(\mathrm{N, N, N})$ is quite stable in time evolution.
For $(\mathrm{N,D,N})$ ,
to reduce ambiguities from the time fluctuation of the energy spectrum, we evaluate the time average of the spectrum. We take the time range as $\tau_0-\Delta \tau\le \tau\le \tau_0+\Delta \tau$ and calculate 
the time-averaged energy spectrum as
\begin{equation}
\bar{e}_n=\frac{1}{2\Delta \tau}\int_{\tau_0-\Delta \tau}^{\tau_0+\Delta \tau} d\tau \left(\frac{E_n}{E}\right).
  \label{energyaverage}
\end{equation}
In this paper, we take $\Delta \tau=4\pi$.

Figure~\ref{figure:energyNDN} shows a direct energy cascade occurs and higher modes are excited at any time.
That is, we can see that the amplitude of some smaller modes decreases with time while the amplitudes of higher modes increase.
We also can find the power-law behavior in the middle region of $1\le n+1\le 40$ as shown in the red points in  figure~\ref{figure:energyNDN}.
Fitting this energy spectra as $\log(\bar{e_n}) = -\alpha \log(n+1)+\beta $ between $1\le n+1\le 40$,
we obtain $\alpha=1.8\pm 0.1 $. 
This fitting curve is plotted in figure~\ref{figure:energyNDN} as a purple solid line.

\subsubsection{Angular momentum}
\label{turbulance:angular_momentum}

By using the coordinate $\sigma'$, we can rewrite the angular momentum $J$ \eqref{conservedchargeEJ} as follows: 
\begin{equation}
    J=\frac{2}{\pi}\int_{-\pi/2}^{\pi/2}d\sigma' \frac{\boldsymbol{\chi}(\sigma(\sigma'))\times p_{\boldsymbol{\chi}}(\sigma(\sigma'))}{\cosh^2(\tan(\sigma'))\cos^2(\sigma')} ,
    \label{JJp}
\end{equation}
where $\eqref{ptptheta,cartesian}$ was used to rewrite $p_\theta$.
Considering the Fourier transformation
\begin{align}
    &\frac{\boldsymbol{\chi}}{\cosh(\tan \sigma')\cos(\sigma')}=\frac{\boldsymbol{\chi}_0}{\sqrt{2}}+\frac{1}{2}\sum_{n=1}^{\infty} (\boldsymbol{\chi}_n e^{2n\sigma'}+\boldsymbol{\chi}_{n}^* e^{-2n\sigma'}) ,\\
    &\frac{p_{\boldsymbol{\chi}}}{\cosh(\tan \sigma')\cos(\sigma')}=\frac{p_{\boldsymbol{\chi}_0}}{\sqrt{2}}+\frac{1}{2}\sum_{n=1}^{\infty} (p_{\boldsymbol{\chi}_n} e^{2n\sigma'}+p_{\boldsymbol{\chi}_{n}}^* e^{-2n\sigma'}) , 
    \label{pchichifourier}
\end{align}
we obtain 
\begin{equation}
  J=\sum_{n=0}^\infty J_n 
  \label{Jsum}
\end{equation}
with 
\begin{equation}
    J_0=\boldsymbol{\chi}_{0}\times p_{{\boldsymbol{\chi}}_{0}},\quad J_n=\frac{1}{2}\left(\boldsymbol{\chi}_n\times p_{\boldsymbol{\chi}_n}^*+\boldsymbol{\chi}_n^*\times p_{\boldsymbol{\chi}_n}\right)=\mathrm{Re}\left(\boldsymbol{\chi}_n\times p_{\boldsymbol{\chi}_n}^*\right) .
    \label{Jn}
\end{equation}

\begin{figure}[t]
  \begin{minipage}[b]{0.45\linewidth}
    \centering
    \includegraphics[keepaspectratio, scale=0.4]{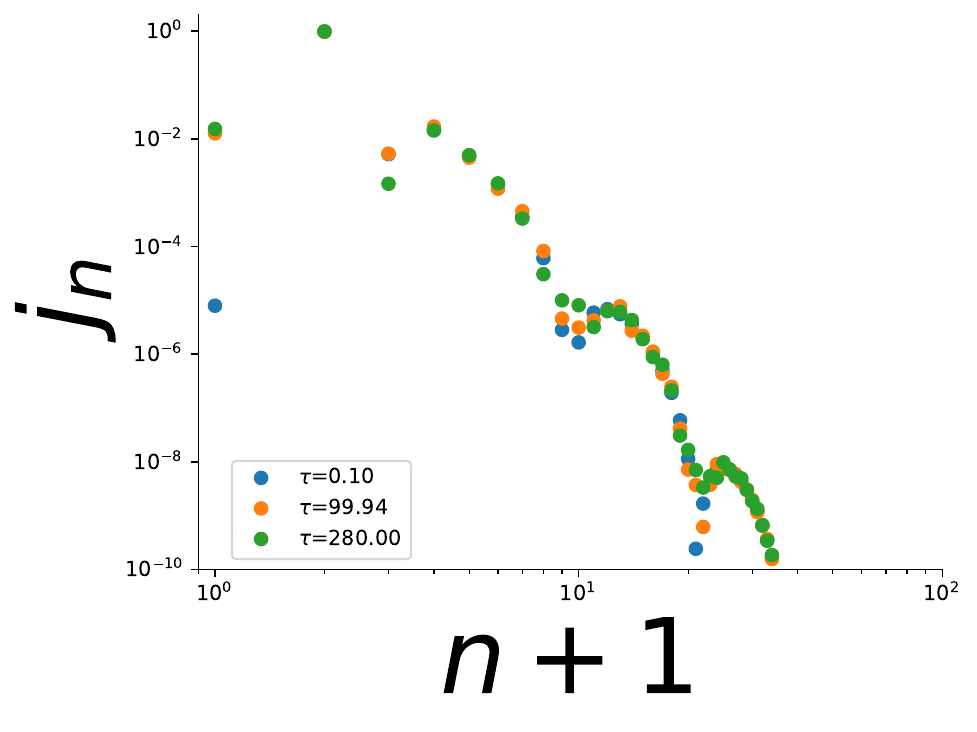}
    \subcaption{$(\mathrm{N, N, N})$}
    \label{figure:angularNNN}
    \end{minipage}
\begin{minipage}[b]{0.45\linewidth}
    \centering
    \includegraphics[keepaspectratio, scale=0.4]{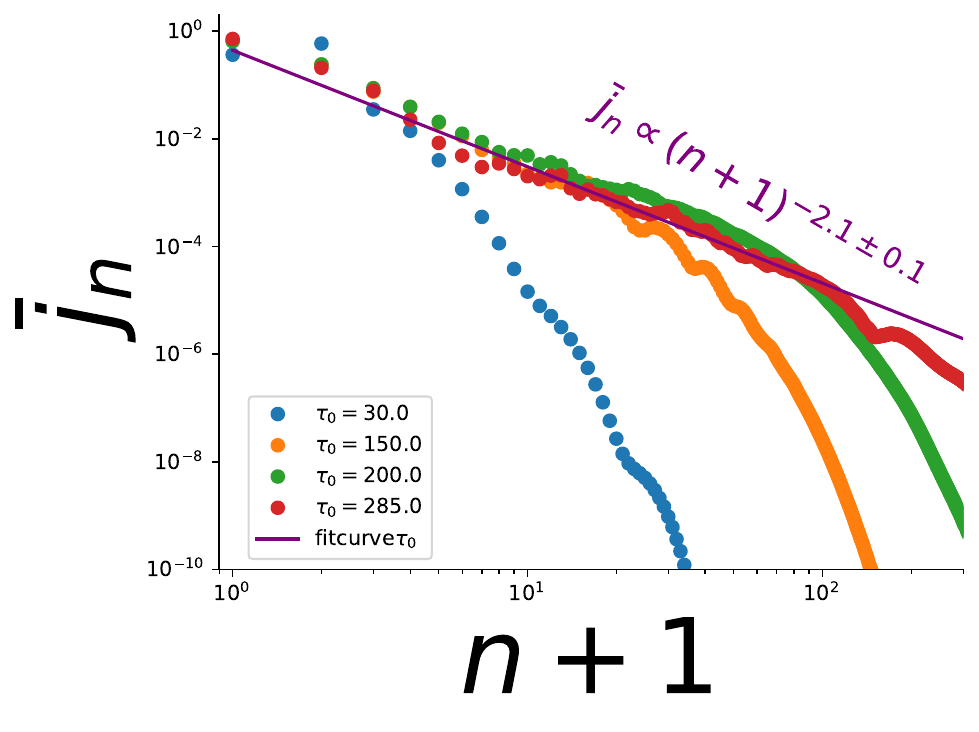}
    \subcaption{$(\mathrm{N, D, N})$}
    \label{figure:angularNDN}
    \end{minipage}
        \caption{
          Angular momentum spectra for  $(\mathrm{N, N, N})$(panel (a)) and $(\mathrm{N,D,N})$ (panel (b)) at each given time. 
          In panel (b), the purple line is a fitting curve for the red points in the range $1\le n+1\le 100$.}
        \label{figure:angular}
\end{figure}

Figure~\ref{figure:angular} shows the angular momentum spectra for 
several values of $\tau$. 
For $(\mathrm{N, N, N})$ (figure~\ref{figure:angularNNN}, where $j_n\equiv |J_n/J|$), 
the angular momentum spectra remain similar to the spectrum in early times.
For $(\mathrm{N}, \mathrm{D}, \mathrm{N})$, 
as in the case of the energy spectrum, 
we take the time average as
\begin{equation}
    \bar{j}_n=\frac{1}{2\Delta \tau}\int_{\tau_0-\Delta \tau}^{\tau_{0}+\Delta \tau} d\tau \left|\frac{J_n}{J}\right| .
         \label{Jnaverage}
 \end{equation}
In figure~\ref{figure:angularNDN} ($\Delta\tau=4\pi$), we can see the excitation of higher modes and the power-law behavior for $\tau_0=285.0$ in the range $1 \le n+1\le 100$. 
Fitting this angular momentum spectrum as $\log \bar{j}_{n} = -\alpha \log(n+1)+\beta $ between $1\le n+1\le 100$, we obtain $\alpha=2.1\pm 0.1 $.
We note that $\bar{j}_1$ decreases in time while $\bar{j}_0$ increases (figure~\ref{figure:angularNDN}). This reflects the dynamics that the angular momentum transfers from spin to orbital angular momenta as we described.

\section{Conclusion}
\label{section:conclusion}

We considered open string dynamics on AdS$_3$ spacetime with the following two boundary conditions. 
$(\mathrm{N, N, N})$: The two endpoints are free, namely, Neumann boundary conditions for all coordinate values.
$(\mathrm{N, D, N})$: The Dirichlet boundary condition is imposed for the radial coordinate $r$ and Neumann boundary conditions for the others $(t,\theta)$. 
Under these boundary conditions, we numerically solved the equations of motion of a perturbed GKP string solution as initial conditions.
For the open string with $(\mathrm{N, N, N})$, which is an integrable boundary condition, the string configuration 
is a stretched fluctuating string that rotates in AdS similar to the reference GKP string solution at any given time as expected. 
In contrast, for $(\mathrm{N, D, N})$, 
we found irregular string motions, where the string crumples in late times. 
We also found a turbulent cascade in 
the energy and angular momentum spectra. 
This result would imply that the open string on AdS$_3$ is non-integrable with the boundary condition $(\mathrm{N, D, N})$ 
while integrable with the boundary condition $(\mathrm{N, N, N})$.
It appears that for $(\mathrm{N, D, N})$, the string is approaching equilibrium state whose typical configuration
is given by a crumpled string configuration with the power-law spectrum. The power law spectrum in such a
typical state have been reported in Refs.~\cite{Bizon:2013xha,Ishii:2016rlk} for systems in (asymptotically) AdS spacetimes.

It would be worthwhile to note that, 
although a turbulent cascade has been observed,
no clear chaotic character was seen in the dynamics of an endpoint of the string (see Appendix~\ref{appendix:chaos}). More investigations would be needed in order to clarify the character of the endpoint dynamics.

It would be also interesting to consider other boundary conditions for the open string in the AdS$_3$. 
For example, $(\mathrm{N, D, D})$ will be another non-integrable boundary condition~\cite{Ishii:2023ucw}. 
Studying turbulent and chaotic behavior in the dynamics of the open string with $(\mathrm{N, D, D})$ would be an interesting future direction.

Since we assume that a ``D-brane'' is placed at a finite radius as in figure \ref{figure:stringmotionmodel}, the holographic interpretation of the turbulent behavior of the open string is unclear. 
What if we put the D-brane to the AdS boundary? Will turbulence still survive in this limit?
The open string in the AdS corresponds to the quark-antiquark pair when string endpoints are put at the AdS boundary. For our  $(\mathrm{N, D, N})$ setup, for example, the endpoints will be smoothly moving on the boundary, dual to moving quarks in the dual field theory.
Looking at the effect of the turbulence in the dual CFT would also be an interesting future work. 

\acknowledgments
The authors would like to thank Dimitrios Giataganas and Kentaroh Yoshida for valuable discussions.
The work of T.I. was supported in part by JSPS KAKENHI Grant Number 19K03871.
The work of R.K.~was financially supported by JST SPRING, Grant Number
JPMJSP2125. R.K.~would like to take this opportunity to thank
``Interdisciplinary Frontier Next-Generation Researcher Program of the Tokai
Higher Education and Research System.''
The work of K.M. was supported in part by JSPS KAKENHI Grant Nos. 20K03976, 21H05186 and 22H01217. 
The work of C.Y. was supported in part by JSPS KAKENHI Grant Nos.~20H05850 and 20H05853.

\appendix 

 \section{Numerical scheme}
 \label{appendix:numericalsheme}
 \subsection{Method of numerical calculations}
 \label{appendix:numericalmethod}
 In this appendix, we explain how to integrate equations of motion~\eqref{stringeom2} numerically.
 We use the numerical method described in \cite{Ishii:2015wua}.
 First, we discretize the double null coordinates $(u,v)$ with the grid spacing $h$ (see figure~\ref{figure:numericalsimulation}). 
 For notational simplicity, we use the unified notation $\phi$ 
 for the fields $t$ and $\boldsymbol{\chi}$ 
 and express $\phi$ at $N$, $E$, $W$, $S$, and $C$ by $\phi_{N}$, $\phi_{E}$, $\phi_{W}$, $\phi_{S}$ and $\phi_{C}$, respectively.  
Around the point $C$, we can approximate $\phi_{,u},\phi_{,v},\phi_{,uv}, \phi$ with a second order accuracy $\mathcal{O}(h^2)$ as 
\begin{equation}
    \phi_{,u}|_{C}=\frac{\phi_{N}-\phi_{E}+\phi_{W}-\phi_{S}}{2h},\quad \phi_{,v}|_{C}=\frac{\phi_{N}-\phi_{W}+\phi_{E}-\phi_{S}}{2h},
    \label{discretization_of_uv}
 \end{equation}   
 and 
 \begin{equation}
    \phi_{,uv}|_{C}=\frac{\phi_{N}-\phi_{E}-\phi_{W}+\phi_{S}}{h^2},\quad \phi|_{C}=\frac{\phi_{W}+\phi_{E}}{2}.
    \label{discretization_of_uvp}
 \end{equation}       
 Substituting these expressions into \eqref{stringeom2}, we obtain nonlinear simultaneous equations.
 Finally, we solve the equations for $\phi_{N}$ with known values $\phi_{E},\phi_{W},\phi_{S}$ as inputs.
 In this paper, we use the Newton-Raphson method for solving the nonlinear simultaneous equations.

 \subsection{Boundary time evolution }
 \label{appendix:boundary_time_evolution}
\begin{figure}[t]
  \centering
  \includegraphics[keepaspectratio, scale=0.6]{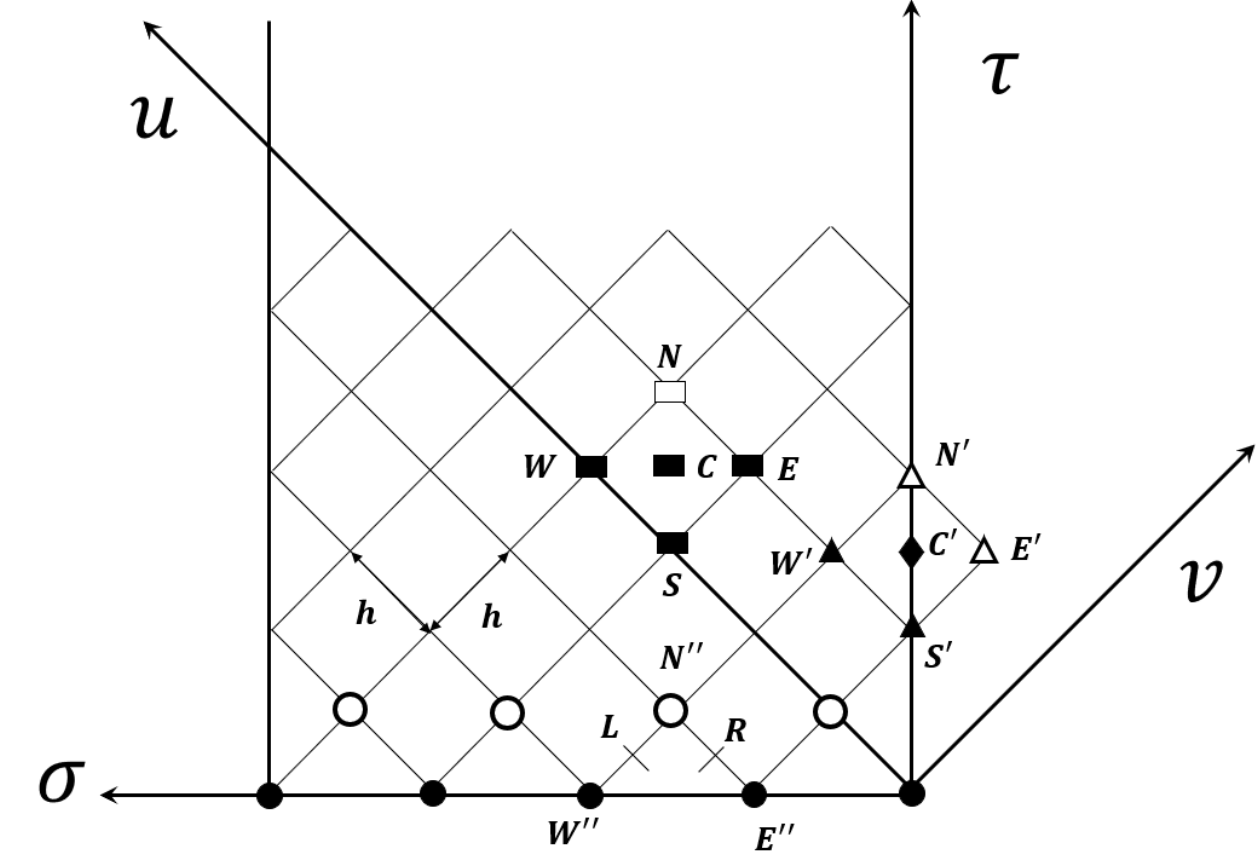}
  \caption{A schematic figure for the numerical scheme.}
  \label{figure:numericalsimulation}
\end{figure}

In this appendix, we explain the numerical scheme for the boundary time evolution.
Figure~\ref{figure:numericalsimulation} shows the worldsheet and the grids for numerical calculations.
For the time evolution, we need to evaluate the value of $(t,\boldsymbol{\chi})$ at $N'$ 
from the known values at $S'$ and $W'$. 
We express $\phi$ at $N'$, $E'$, $W'$, $S'$, and $C'$ by $\phi_{N'}$, $\phi_{E'}$, $\phi_{W'}$, $\phi_{S'}$ and $\phi_{C'}$, respectively. 

The general approach is as follows. 
First, we obtain the value of $\phi $ with imposed Neumann boundary condition at the ghost point $E'$ by using the boundary condition at $C'$.
To do this, we approximate the value of $\phi_{,\sigma}$ as 
\begin{equation}
    \phi_{,\sigma}|_{C'}=\frac{\phi_{W'}-\phi_{E'}}{\sqrt{2}h} +\mathcal{O}(h^2).
    \label{phisigma}
  \end{equation}
  By substituting this Eq.~\eqref{phisigma} into the boundary condition, we obtain $\phi$ at $E'$.
Second, we obtain $\boldsymbol{\chi}$ at the point where $\boldsymbol{\chi}$ is not determined by boundary condition, using EOM \eqref{stringeom2} with boundary conditions:
for $\boldsymbol{\chi}$ with Neumann boundary conditions, we obtain $\boldsymbol{\chi}$ at $N'$ using the EOM, while  
for $\boldsymbol{\chi}$ with Dirichlet boundary conditions, we obtain $\boldsymbol{\chi}$ at $E'$ using the EOM.
Finally, we derive $t$ at $N'$ by using constraints.
We will see the more specific procedures for the boundary conditions $(\mathrm{N, N, N})$ and $(\mathrm{N, D, N})$ in the following subsections.

\subsubsection{$(\mathrm{N, N, N})$}
The boundary conditions at $C'$ are given by 
\begin{equation}
    t_{,\sigma}|_{C'}=0,\quad \boldsymbol{\chi}_{,\sigma}|_{C'}=\boldsymbol{0}.
\end{equation}
Then we can obtain $t_{E'}$ and $\boldsymbol{\chi}_{E'}$ as follows: 
\begin{equation}
    t_{E'}=t_{W'},\quad \boldsymbol{\chi}_{E'}=\boldsymbol{\chi}_{W'}.
    \label{NNNcoordinate}
\end{equation}
We obtain $\boldsymbol{\chi}_{N'}$ by the equation of motion with~\eqref{NNNcoordinate}.
Then we obtain $t_{N'}$ by using the constraint 
\begin{equation}
    t_{N'}=t_{S'}+\frac{2}{(1+|\boldsymbol{\chi}_{W'}|^2)}|\boldsymbol{\chi}_{N'}-\boldsymbol{\chi}_{S'}|.
\end{equation}

\subsubsection{$(\mathrm{N, D, N})$}
For simplicity, we work with the polar coordinate system given by Eq.~\eqref{AdS3 metric}.
We can derive $r_E'$, $r_S'$, $\theta_E'$ and $\theta_S'$ by the coordinate transformation~\eqref{poloarcoordinate}.
In the case of $(\mathrm{N, D, N})$, $r$ at the boundary is fixed to the constant value $r_0$, and the boundary conditions at $C$ are
\begin{equation}
    t_{,\sigma}|_{C'}=0,\quad r_{C'}=r_0,\quad \theta_{,\sigma}|_{C'}=0 .
    \label{NDN}
\end{equation}
From this, we obtain 
\begin{equation}
    t_{E'}=t_{W'},\quad \theta_{E'}=\theta_{W'},\quad r_{N'}=r_{S'}.
       \label{boundaryNDN} 
\end{equation}
From  $\theta_{,uv}|_{C'}=0$, $\theta_{N'}$ is obtained by 
\begin{equation}
    \theta_{N'}=2\theta_{W'}-\theta_{S'} .
    \label{NDNtheta}
\end{equation}
In order to find the value of $r_{E'}$, we use EOM \eqref{stringeom2} for $r$. Then 
we obtain $t_{N'}$ by using the constraint: 
\begin{equation}
    t_{N'}=t_{S'}+\frac{\sqrt{(r_{W'}-r_{E'})^2+r_{0}^2(\theta_{N'}-\theta_{S'})^2}}{(1+r_0^2)} .
\end{equation}
By the coordinate transformation to the Cartesian coordinates $\boldsymbol{\chi}$, we can obtain $\boldsymbol{\chi}_{N'}$.

 \subsection{Construction of initial data}
 \label{appendix:constructionginitialvalue}
In this appendix, we explain how to prepare initial data based on \cite{Ishii:2016rlk}.
See figure~\ref{figure:numericalsimulation}.
Configuration of $(t,\boldsymbol{\chi})$ on $\tau=0$ ($\bullet$) is determined by Eq.~\eqref{initial value}.

Then on the next surface $\tau=\Delta \tau $ ($\circ$) we give a configuration of $\chi_2$ by Eq.~\eqref{initial value}.
For  $t$ and $\chi_1$ on $\tau=\Delta \tau $, we determine them by using constraints.  
Let us consider how to determine $t_N$ and  $\boldsymbol{\chi}_N$.
First, we define the point $L$($R$) as the middle point between $N$ and $E$($N$ and $W$).
Then we approximate $\phi$ and the derivative of $\phi$ as follows:
\begin{equation}
    \phi_{,u}|_{R}=\frac{\phi_N-\phi_E}{h}+\mathcal{O}(h^2),\quad  \phi_{,u}|_{L}=\frac{\phi_N-\phi_W}{h}+\mathcal{O}(h^2) ,
    \label{appendix:initialvalue}
\end{equation}
 and 
\begin{equation}
    \phi_R=\frac{\phi_N+\phi_E}{2}+\mathcal{O}(h^2),\quad \phi_L=\frac{\phi_N+\phi_W}{2}+\mathcal{O}(h^2) .
\end{equation}
Since $h_{uu}=0$, $t_E=0$ and $t_N>0$, from Eq.~\eqref{induced metric explicit} on $R$, we obtain 
\begin{equation}
    t_N=2\frac{\sqrt{(\chi_{1N}-\chi_{1E})^2+(\chi_{2N}-\chi_{2E})^2}}{1+\frac{(\chi_{1N}+\chi_{1E})^2}{4}+\frac{(\chi_{2N}+\chi_{2E})^2}{4}} .
    \label{appendix:initialvaluehuu}
\end{equation}
Similarly to the point $R$, since $h_{vv}=0$, $t_W=0$ and $t_N>0$ on the point $L$, we obtain  
\begin{equation}
    t_N=2\frac{\sqrt{(\chi_{1N}-\chi_{1W})^2+(\chi_{2N}-\chi_{2W})^2}}{1+\frac{(\chi_{1N}+\chi_{1W})^2}{4}+\frac{(\chi_{2N}+\chi_{2W})^2}{4}} .
    \label{appendix:initialvaluehvv}
\end{equation}
Eliminating $t_N$ from Eqs.~\eqref{appendix:initialvaluehvv} and \eqref{appendix:initialvaluehuu}, we get the following equation:
\begin{equation}
    \frac{\sqrt{(\chi_{1N}-\chi_{1E})^2+(\chi_{2N}-\chi_{2E})^2}}{4+(\chi_{1N}+\chi_{1E})^2+(\chi_{2N}+\chi_{2E})^2}=\frac{\sqrt{(\chi_{1N}-\chi_{1W})^2+(\chi_{2N}-\chi_{2W})^2}}{4+(\chi_{1N}+\chi_{1W})^2+(\chi_{2N}+\chi_{2W})^2} . 
    \label{appendix:intialvaluechi1}
\end{equation}
Since the value of $\chi_{2N}$ is already known from Eq.~\eqref{initial value}, Eq.~\eqref{appendix:intialvaluechi1} can be regarded as an equation for $\chi_{1N}$. 
Then we can solve Eq.~\eqref{appendix:intialvaluechi1} for $\chi_{1N}$ by using the Newton method. 
The value of $t_N$ can be given by substituting $\chi_{1N}$ into Eq.~\eqref{appendix:initialvaluehuu} or Eq.~\eqref{appendix:initialvaluehvv}. 

 \section{Error analysis}
 \label{appendix:Error analysis}
 
 In this appendix, we evaluate numerical errors due to discretization as a constraint violation.
 Defining  $C_u$ and $C_v$ as
 \begin{equation}
    C_u(\,\sigma)= (1+|\boldsymbol{\chi}|^2)^2t_{,u}^2-{4|\boldsymbol{\chi}_u|^2},\quad C_v(\tau,\sigma) = (1+|\boldsymbol{\chi}^2|^2)t_{,v}^2-{4|\boldsymbol{\chi}_v|^2}, 
\end{equation}
we can write the constraints~\eqref{Virasoro} as $C_u=C_v=0$. 

Then we define $C_{max}$ as 
\begin{equation}
    C_{\mathrm{max}}(\tau)=\underset{-\frac{\pi}{2}\le\sigma\le\frac{\pi}{2}}{\max}\frac{|C_u|+|C_v|}{2((1+|\boldsymbol{\chi}|^2)t_{,u}^2+{4|\boldsymbol{\chi}_u|^2}+(1+|\boldsymbol{\chi}|^2)t_{,v}^2+{4|\boldsymbol{\chi}_v|^2})}.
    \label{appendix:Cmax}
\end{equation}

\begin{figure}[t]
    \centering
    
    \includegraphics[keepaspectratio, scale=0.4]{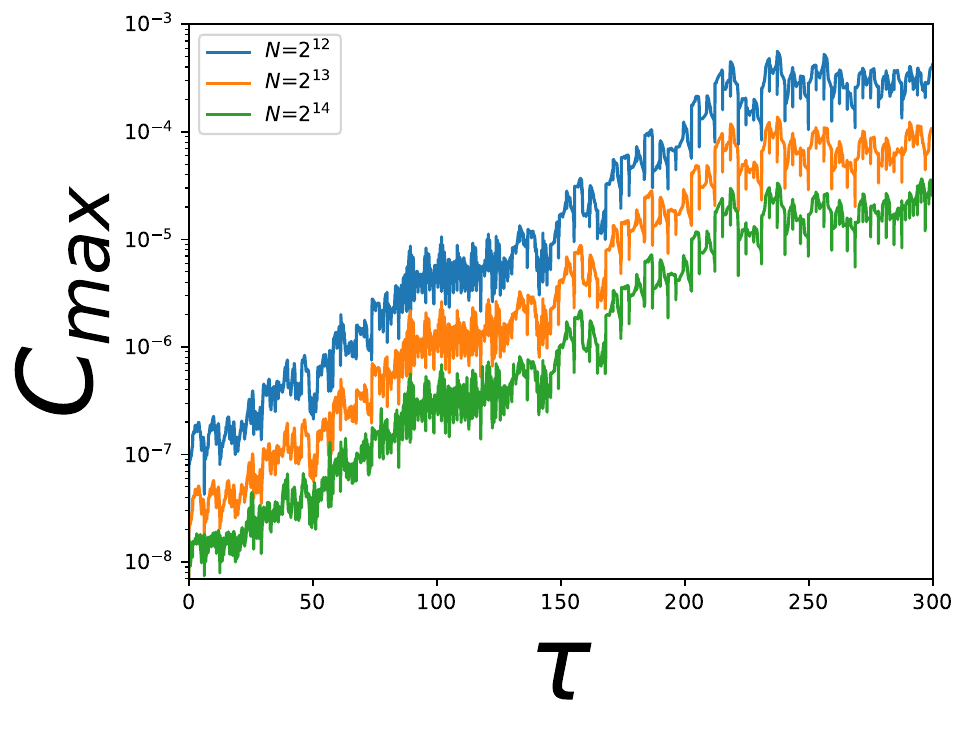}
    \caption{Constraint violation for  $(\mathrm{N, D, N})$ as functions of $\tau$ for $N=2^{12}$, $2^{13}$ and $2^{14}$, where $N$ is the number of grids.  }
    \label{figure:error}
\end{figure}

In figure~\ref{figure:error}, $C_{\mathrm{max}}$ for $(\mathrm{N, D, N})$ is plotted for several grid numbers $N$.
We can find $C_{\mathrm{max}}\propto 1/N^2$ which is consistent with our second order discretization scheme.   

\section{Conserved charges}
\label{appendix:conserved charge}
In this appendix, we explicitly 
show that
the conserved charges defined for $( \mathrm{N, N, N})$ are not conserved for $(\mathrm{N, D, N})$.

For (N, N, N), there are two conserved quantities $ M_{a,b}$  which are obtained by taking the trace of the monodromy matrices \cite{Ishii:2023ucw}. These are equipped with the spectral parameter $\lambda$, and by Taylor expanding them with respect to $\lambda$, 
an infinite number of conserved quantities are obtained as the coefficients in the expansion. At $\mathcal{O}(1/\lambda^2)$ in the expansion around $\lambda\to\infty$, we have
\begin{equation}
M_{a}^{(2)}=2\int_{-\pi/2}^{\pi/2}d\sigma\int_{-\pi/2}^{\pi/2}d\sigma'J_{\tau A}(\tau,\sigma)J_{\tau}^{A}(\tau,\sigma'),\quad M_{b}^{(2)}=2\int_{-\pi/2}^{\pi/2}d\sigma\int_{-\pi/2}^{\pi/2}d\sigma'I_{\tau A}(\tau,\sigma)I_{\tau}^{A}(\tau,\sigma'), 
\label{appendix:conservedJI}
\end{equation}
where $J_{tA}, I_{tA}$ are conserved currents associated with $SL(2,R)\times SL(2,R)$ symmetries and $A$ runs over 1 to 3.
We raise the index $A$ by matrix $\gamma^{AB}=\frac{1}{2}\mathrm{diag}(-1,1,1)$ and lower by the inverse of $\gamma^{AB}$: $\gamma_{AB}=2\mathrm{diag}(-1,1,1)$.
The components of $J_{\tau A}, I_{\tau A}$ are explicitly given by 
\begin{eqnarray}
    \begin{aligned}
    &J_{\tau0}=\frac{-(1+r^2)^2t_{,\tau}+4r^2\theta_{,\tau}}{2(1-r^2)},\\
    &J_{\tau 1}+i J_{\tau 2}=\frac{e^{-i(t+\theta)}(r(1+r^2)(t_{,\tau}-\theta_{,\tau})-i (1-r^2)r_{,\tau})}{(1-r^2)^2},
    \end{aligned}
        \label{appendix:J expilicit}
    \end{eqnarray}
and 
\begin{eqnarray}
    \begin{aligned}
        &I_{\tau0}=\frac{(1+r^2)^2t_{,\tau}+4r^2\theta_{,\tau}}{2(1-r^2)},  \\
        &I_{\tau1}+i I_{\tau2}=\frac{e^{i(t-\theta)}(-i r(1+r^2)(t_{,\tau}+\theta_{,\tau})+ (1-r^2)r_{,\tau})}{(1-r^2)^2}.      
    \end{aligned}
    \label{string:I expilicit}
\end{eqnarray}
  
\begin{figure}[t]
    \centering
    \begin{minipage}[b]{0.45\linewidth}
        \centering
        \includegraphics[keepaspectratio, scale=0.4]{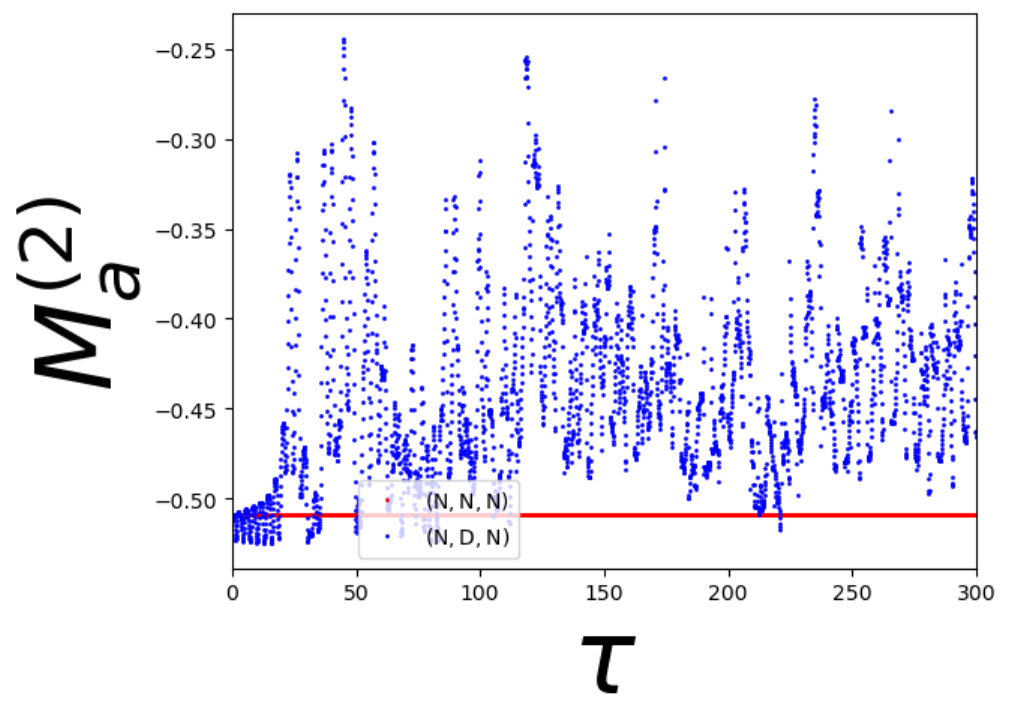}
        \subcaption{$M_{a}^{(2)}$}
        \label{figure:conservedJ}
        \end{minipage}
    \begin{minipage}[b]{0.45\linewidth}
    \centering
    \includegraphics[keepaspectratio, scale=0.4]{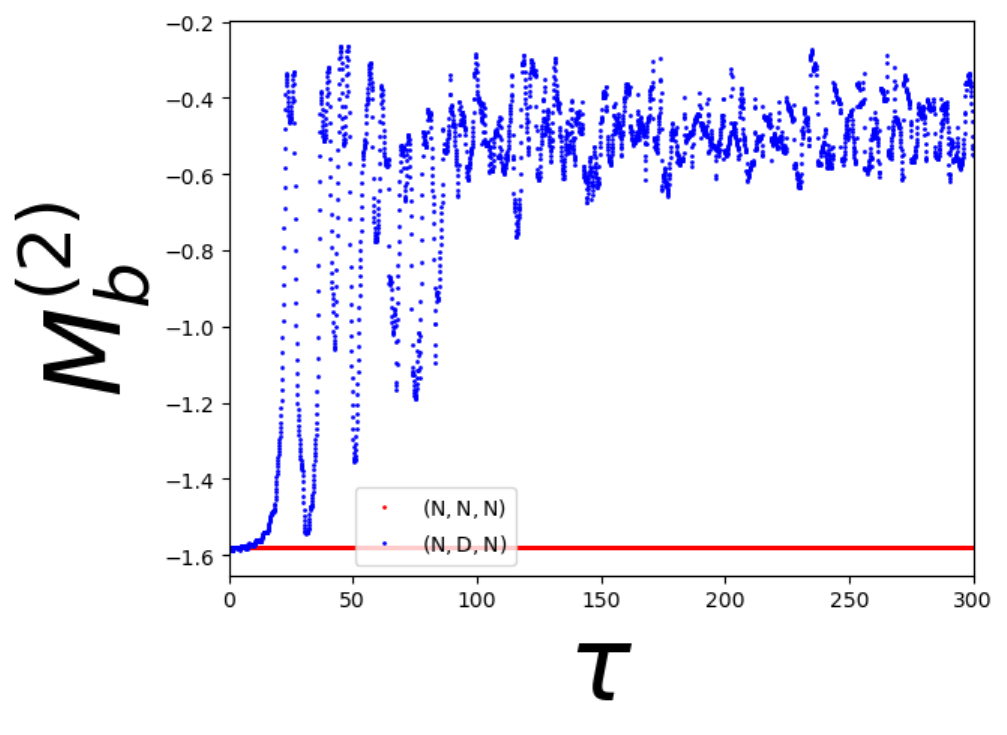}
    \subcaption{$M_{b}^{(2)}$}
    \label{figure:conservedI}
    \end{minipage}    
    \caption{The time dependence of $M_{a}^{(2)}$(panel$(\mathrm{a})$), $M_{b}^{(2)}$(panel$(\mathrm{b})$) for $(\mathrm{N, N, N})$ (red) and $(\mathrm{N, D, N})$ (blue).}
    \label{figure:stringconserved}
\end{figure}

Figure~\ref{figure:stringconserved} shows the time dependence of  $M_{a}^{(2)},M_{b}^{(2)}$ for $(\mathrm{N, N, N})$ and $(\mathrm{N, D, N})$. 
As we can see, while both $M_{a}^{(2)}$ and $M_{b}^{(2)}$ do not change in time for $(\mathrm{N, N, N} )$, they change for $(\mathrm{N, D, N} )$.
Thus the conserved charges constructed for $(\mathrm{N, N, N})$ are not
conserved for $(\mathrm{N, D, N})$. 

\section{Analysis of the sensitivity to initial conditions for $(\mathrm{N, D, N})$}
\label{appendix:chaos}
In this appendix, we give the analysis of sensitivity to initial conditions for $(\mathrm{N, D, N})$.
Let us check the sensitivity of the string configuration to the parameter $\epsilon$. 
In the following, we focus on the (non-)integrability on the string worldsheet due to the boundary conditions, and hence we will perform the analysis with the time coordinate $\tau$.
To quantitatively evaluate the sensitivity to the initial conditions, we introduce the Lyapunov exponent $\lambda$ for the angular location of the string endpoints as follows.
In numerical calculations, practically, the domain of $\theta(\tau,\sigma)$ is defined in $-\infty<\theta<\infty$, where the angular coordinate of the GKP string at the initial time $\tau=0$ is $\theta=0,\pi$ (see figure~\ref{figure:stringmotionNNN}). Such $\theta$ takes into account how many times the string rotates along the time evolution.
For example, $\theta(\tau,\frac{\pi}{2}) = 2\pi n$ with $n$ being integers implies that the endpoint has rounded $n$ times anti-clockwise ($|n|$ times clockwise if $n<0$).
Then, we introduce the Lyapunov exponent $\lambda$ by \cite{ott_2002}
\begin{equation}
  \lambda =\lim_{\tau \to \infty} \lim_{\delta\epsilon \to 0} \frac{1}{\tau}\log \frac{|\delta \theta_{-\pi/2}(\tau,\delta\epsilon)|}{|\delta \theta_{-\pi/2}(0,\delta \epsilon)|},
  \label{Lyapunov exponent}
  \end{equation} 
where we introduced the difference of angular coordinates of string endpoints as
\begin{equation}
  \delta \theta_{-\pi/2} (\tau,{\delta\epsilon}) =\Bigg| \theta\left(\tau,-\frac{\pi}{2}\right)\bigg|_{\epsilon=0.1+\delta \epsilon}-\theta\left(\tau,-\frac{\pi}{2}\right)\bigg|_{\epsilon=0.1}\Bigg|.
  \label{dtheta}
\end{equation}

\begin{figure}[t]
    \centering
    \includegraphics[keepaspectratio, scale=0.6]{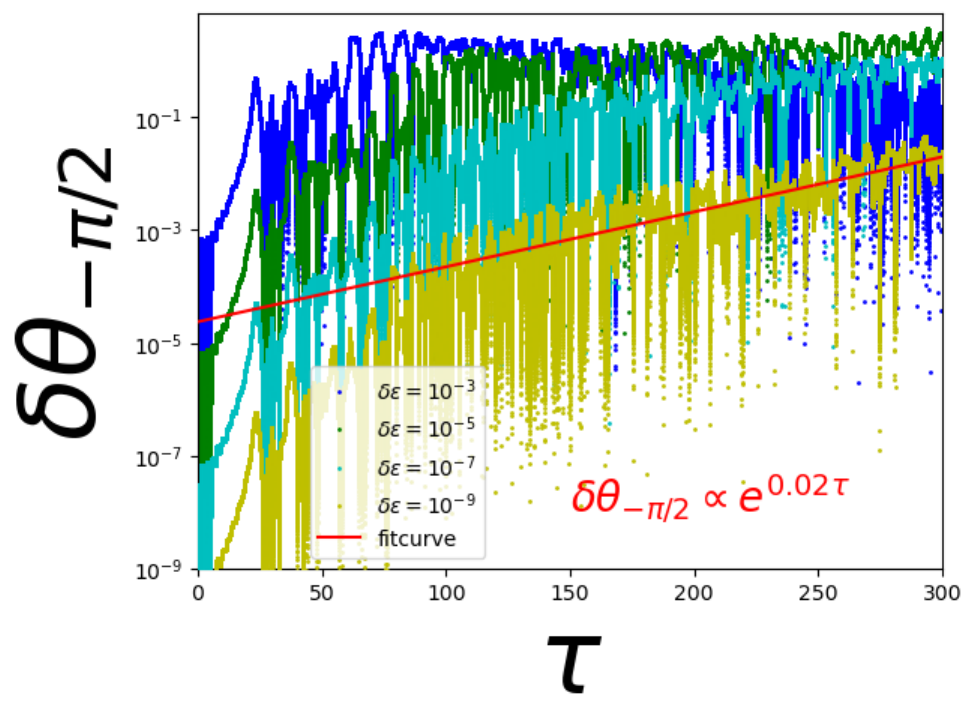}
    \caption{The values of $\delta \theta_{-\pi/2} $ as functions of $\tau$ for $\delta\epsilon =10^{-3}$, $10^{-5}$, $10^{-7}$,$10^{-9}$ for $(\mathrm{N,D,N})$. The time range for the fit is $150\le\tau\le300$. }
    \label{figure:dthetaNDN}
 \end{figure}

\begin{figure}[t]
    \centering
    \includegraphics[keepaspectratio, scale=0.6]{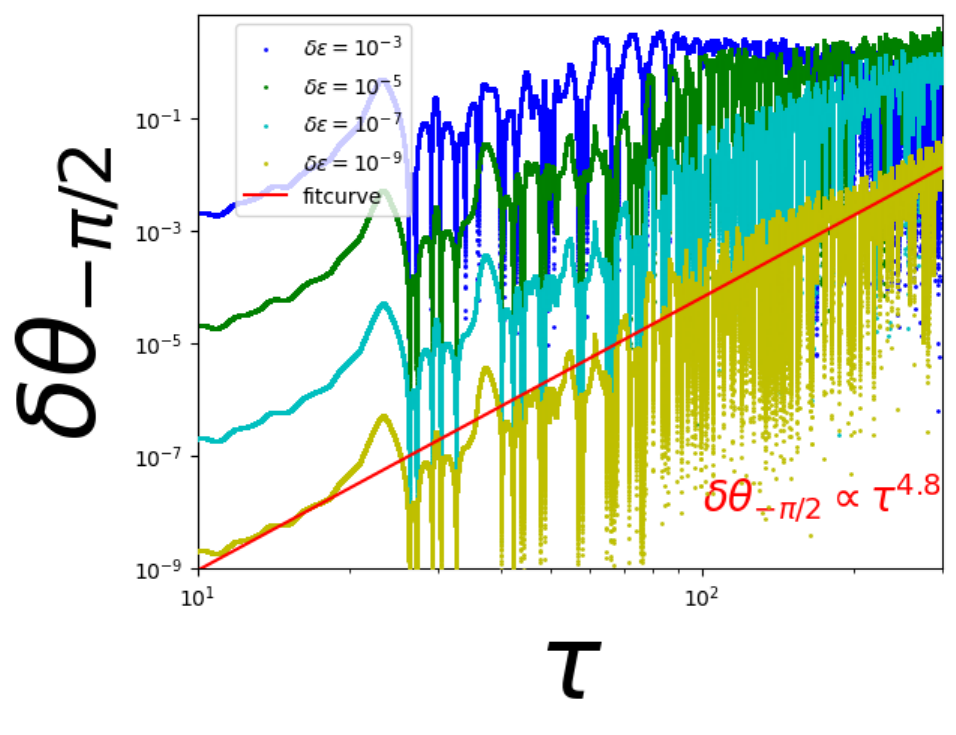}
    \caption{The values of $\delta \theta_{-\pi/2} $ as functions of $\tau$ for $\delta\epsilon =10^{-3}$, $10^{-5}$, $10^{-7}$,$10^{-9}$ for $(\mathrm{N,D,N})$. The data is the same as figure~\ref{figure:dthetaNDN} but shown in a log-log plot. The time range for the fit is $150\le\tau\le300$. }
    \label{figure:dthetaNDNpower}
  \end{figure}

Figure~\ref{figure:dthetaNDN} shows the time evolution of $\delta \theta_{-\pi/2} (\tau,{\delta\epsilon})$ for $\delta\epsilon=10^{-3},10^{-5},10^{-7},10^{-9}$ for $(\mathrm{N,D,N})$.
To find sensitivity to the conditions, we should consider the late time behavior 
where the asymptotic behavior would dominate in 
\eqref{Lyapunov exponent}.
In this appendix, we consider the time range $150\le\tau\le300$.
In this time range, however, we could not distinguish power and exponential growth clearly (figures~\ref{figure:dthetaNDN} and \ref{figure:dthetaNDNpower}). 
If we estimate the Lyapunov exponent by fitting  $\delta \theta_{-\pi/2} (\tau,{\delta\epsilon})=A \exp (\lambda \tau )$ to the data for $\delta \epsilon=10^{-9}$ ,
the Lyapunov exponent $\lambda$ is given by $\lambda\simeq 0.02$. Instead, if we fit by a power law function $\delta \theta_{-\pi/2} (\tau,{\delta\epsilon})=B \tau^{\alpha} $, we get $\alpha \simeq$4.8 (figure~\ref{figure:dthetaNDNpower}). The power law fitting appears to explain the behavior of $\delta \theta$ better even in earlier times than the exponential one.

\section{Numerical checks for the conservations of $E$ and $J$}
\label{appendix:numerical check of E J}
\begin{figure}[t]
    \centering
    \begin{minipage}[b]{0.45\linewidth}
        \centering
        \includegraphics[keepaspectratio, scale=0.4]{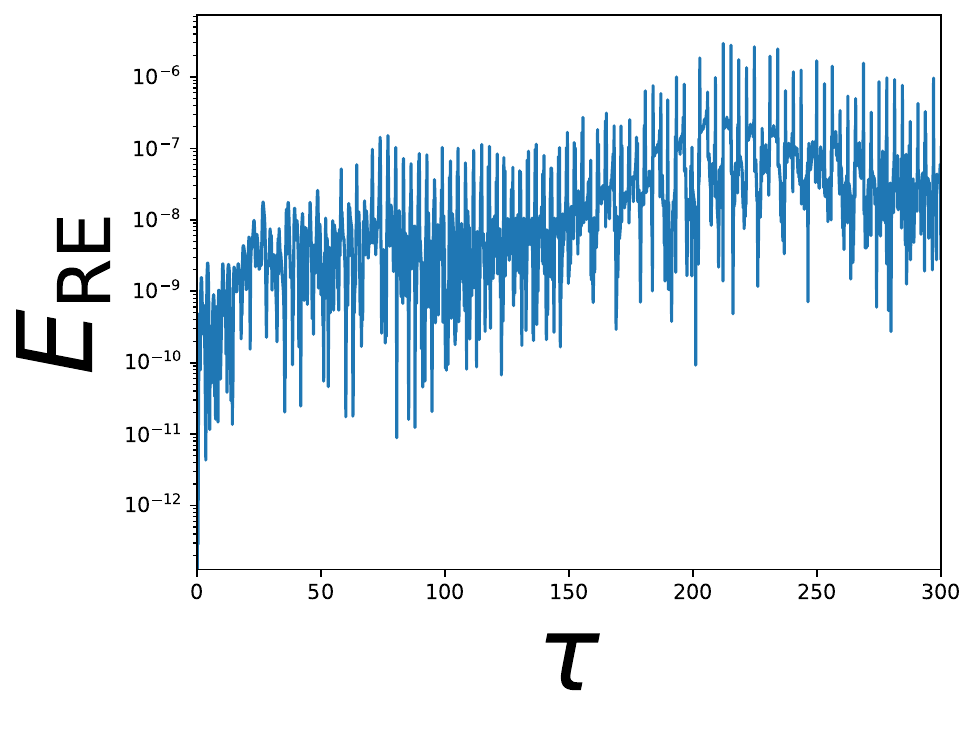}
        \subcaption{$E_{\mathrm{RE}}$}
        \label{figure:conserveE}
        \end{minipage}
    \begin{minipage}[b]{0.45\linewidth}
    \centering
    \includegraphics[keepaspectratio, scale=0.4]{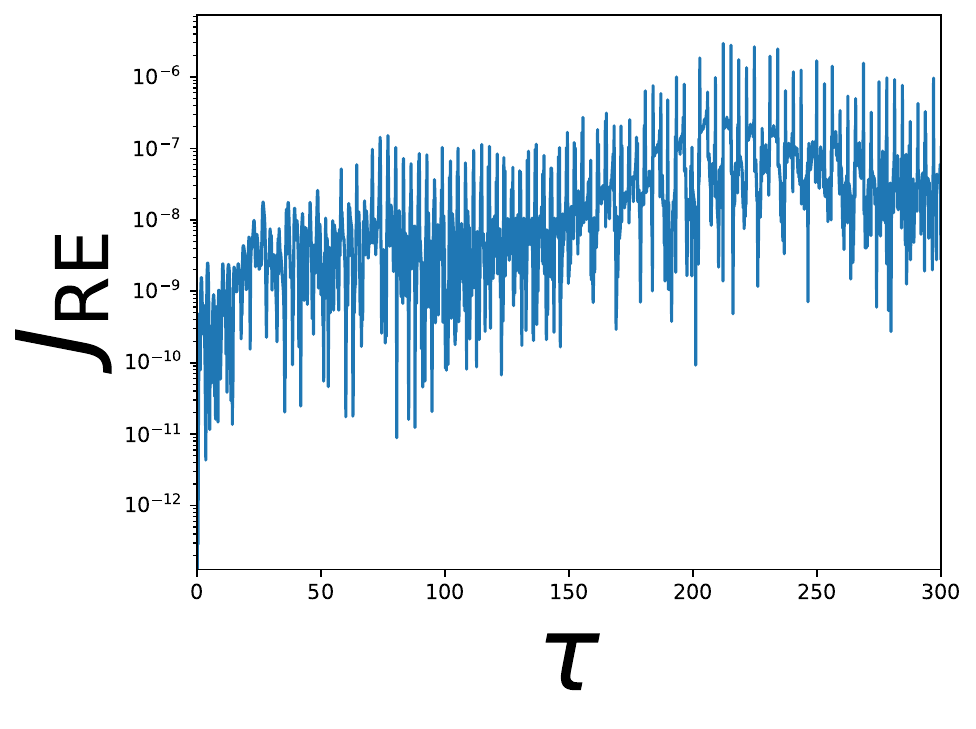}
    \subcaption{$J_{\mathrm{RE}}$}
    \label{figure:conserveJ}
    \end{minipage}    
    \caption{The time dependence of $E_{\mathrm{RE}}$(panel$(\mathrm{a})$), $J_{\mathrm{RE}}$(panel$(\mathrm{b})$) for $(\mathrm{N, D, N})$.}
    \label{figure:stringconservedEandJ}
\end{figure}

In this appendix, we check that the energy $E$~\eqref{Esigmap} and angular momentum $J$~\eqref{JJp} are conserved for $(\mathrm{N, D,N})$.
To evaluate the numerical violation of the conservation, we define the relative errors $E_{\mathrm{RE}}, J_{\mathrm{RE}}$ as follows:
\begin{equation}
 E_{\mathrm{RE}}=\left|\frac{E(\tau)-E(0)}{E(0)}\right|, \quad   J_{\mathrm{RE}}=\left|\frac{J(\tau)-J(0)}{J(0)}\right|.
 \label{relative_error E and J}
\end{equation}
In figure \ref{figure:stringconservedEandJ}, we can see the violations are no more than numerical errors (see figure~\ref{figure:error}).
This confirms that $E$ and $J$ are conserved within numerical errors\footnote{Numerical errors in a single mode the in energy spectrum is smaller than $E_{RE}$ and that the errors are not the reason for the power law behavior for $(\mathrm{N, D, N})$ in figure \ref{figure:energyNDN}.}.

 \bibliography{chaos_and_turbulence_of_open_string}
 \bibliographystyle{JHEP}

\end{document}